\documentclass{aa}  
\usepackage{graphicx}
\usepackage{txfonts}
\usepackage{natbib} 

\usepackage{hyperref}
\begin{document}

\title{The outskirts of M33: Tidally induced distortions versus signatures of gas accretion }

\titlerunning{No tidally induced distorsions in M33 }  
 
\author{Edvige Corbelli
       \inst{1}
        \and 
           Andreas Burkert
       \inst{2}
         }

   \institute{INAF-Osservatorio Astrofisico di Arcetri, Largo E. Fermi, 5,
             50125 Firenze, Italy\\
             \email{edvige.corbelli@inaf.it} 
             \and 
             Universit\"as-Sternwarte, Ludwig-Maximilians-Universit\"at M\"unchen, Scheinerstr. 1, D-81679 Munich, Germany\\
                   \email{burkert@usm.uni-muenchen.de}
                    }                                   
 
   \date{Received .....; accepted ....}

 \abstract
   {  }
   {We investigate a possible close encounter between M33 and M31 in the past to understand the role of galaxy-galaxy interactions  in shaping  the matter distribution in galaxy outskirts. }
   { 
   By examining a variety of initial conditions, 
   we recovered possible orbital trajectories of M33, M31 and the Milky Way in the past, which are compatible with the Early Third Data Release of the {\it Gaia} mission and with mass estimates of Local Group spirals. 
 Using test-particle  simulations we explored if the M33 warp and its dark satellite distribution  have been induced by a past M33-M31 encounter  along these orbits, after tuning mass losses and the dynamical friction term with the help of  N-body numerical simulations.  
   }
   {  A close encounter  of M33 and M31 in the past has a low   but non-negligible probability. If the two galaxies had been closer in the past, their minimum distance would be of the order of 100~kpc or larger, and this happened earlier than 3~Gyr ago. During this encounter, 35-40$\%$ of the dark matter mass  of M33 might have been removed from the halo  due to tidal stripping.
A detailed comparison of the results of test-particle simulations with  the observed disk warp  or with the  spatial distribution of candidate dark satellites  of M33  suggests that a closer passage of M33 around M31 cannot, however, be responsible for the observed morphological features.   We suggest that more recent gas accretion events, possibly from a cosmic filament,  might cause the misalignment of the outer disk of M33 after the rapid  inner disk formation. 
  }
  {}

   \maketitle
 
\section{Introduction}

The relevance of galaxy outskirts for understanding galaxy formation and evolution is fully recognised today 
because numerical simulations and observations show that galaxies do not evolve as close boxes but
through a continuous exchange of radiation and matter with the environment,  cosmic filaments, 
clusters, and groups where  galaxies reside \citep{2006MNRAS.368....2D,
2006ApJ...650..560C,2009ApJ...700L...1K,2009AN....330..904B}.   
Depending upon cosmic time, the cosmic web  as well as galaxy encounters might trigger or stop gas accretion,   
driving internal processes such as the rate of star formation and galaxy growth \citep{1987AJ.....93.1011K,2002MNRAS.332..155L,2014MNRAS.444.2071D}.  
Evidence of gas accretion from the cosmic web, today or at earlier cosmic times, is still 
controversial since direct imaging  of cosmic filament flows is difficult due to the low matter density and to projection effects.   
There is only indirect evidence that this is effectively taking place \citep{2014A&ARv..22...71S,2015ApJ...812...58C}. 
Galaxy outskirts are excellent places where one can search for signs of recent gas accretion or tidal disturbances because of their
low baryonic content and shallower potential well than the brighter inner regions.

Local Group galaxies offer a unique opportunity to search for signatures of these processes because of the high spatial resolution and sensitivity available \citep{2004A&A...417..421B,2004NewAR..48.1271B,2008A&A...487..161G,2009ApJ...703.1486P,2013Natur.497..224W,2016A&A...589A.120K,
2016ApJ...816...81W}. 
In the Local Group, however, the evidence of intergalactic cold gas accretion into its more massive members, 
the Milky Way (hereafter MW), M31, and M33, has to face another difficulty. Internal cycling such as galactic fountains or cosmic web accretion can produce cold gas clouds 
above the MW disk, also known as high velocity clouds (HVCs), with systemic velocities close to that of M31 or M33. This makes it difficult to distinguish 
MW clouds lying in projection towards M31 or M33 from more massive and distant clouds effectively orbiting these galaxies.
Gas clouds that show a rotational pattern are likely to be associated with M31 or M33  \citep{2004ApJ...601L..39T,2005A&A...436..101W,2008A&A...487..161G}.  
These clouds at the distance of M31 and M33 can be a sign of gas accretion or the baryonic counterparts of dark satellites orbiting M31 or M33, alleviating the missing satellite problem, which is very severe  in the case of  M33  \citep{2011A&A...533A..91G,2016MNRAS.456..951K,2018MNRAS.480.1883P,2022MNRAS.509...16M}.
 
The outer disk of M33 has a well-defined orientation established through a tilted ring model fit to the velocity field of the 
21-cm line \citep{1997ApJ...479..244C,2014A&A...572A..23C}. The presence of a warp in the outer disk can be a sign of a recent interaction 
but also of slow gas and dark matter accretion at later times than the inner disk formation epoch, which causes an infall-driven reorientation 
of the outer parts of virialised haloes and disks \citep{1999MNRAS.303L...7J}.
In this paper we examine, in detail, the possible origins of the disk warp and of the non-isotropic distribution of possible dark satellites, as traced by HI gas clouds in the proximity of M33 \citep{2008A&A...487..161G,2014A&A...572A..23C}, focussing on simulating the possible past encounter between M31 and M33 as the origin of these morphological features.
The gaseous mass of   HI clouds in the outskirts of M33 varies between 10$^5$ and 10$^6$~M$_\odot$ considering also the ionised gas. If they are due to
a steady cosmic flow, where the gas compresses and recombines as it enters the potential well of M33,  
\citet{2008A&A...487..161G}  estimated that a likely mass flow rate would be of order of 1~M$_\odot$~yr$^{-1}$. This is 
similar to the star formation rate of M33 \citep{2009A&A...493..453V} , that is to say what flows in is converted into stars.
Numerical results of cosmological simulations do indeed show that galaxies in groups with a dark matter mass similar to that of M33 are not 
too massive to stop cold streams via shock heating,  and today they  can still accrete gas from the cosmic web  along preferential planes \citep{2009ApJ...700L...1K,2009ApJ...694..396B}.
The discovery of a satellite plane around M31 can also be another piece of evidence of the alignment of local structures with the 
cosmic environment and casts serious doubts on a past interaction between M33 and M31 \citep{2013MNRAS.436.2096S}. 

On the other hand, if HI clouds are the visible counterparts of low-mass galaxies, their spatial distribution can be non-homogeneous because of a past interaction between M31 and M33.
The presence of stellar streams around M31 \citep{2006ApJ...647L..25M,2007ApJ...671.1591I,2010ApJ...723.1038M,
2013ApJ...763....4L}, some of which point in the direction of M33, has also triggered numerous studies on their possible tidal origins.
Although most of the streams around M31 may be due to disruption of low-mass satellites, it is possible that a close
passage of M33 around M31 in the past  has stripped gas and stars from the outskirts of these two spirals
\citep{2008MNRAS.390L..24B,2009Natur.461...66M}. 

 The question of a close encounter in the past between M31 and M33 is still controversial. 
 The measure of the transverse motion of Local Group galaxies has recently become possible thanks to radio interferometers such as VLBI (Very Long Baseline Interferometry), to Hubble Space Telescope (HST) data 
 and  to the release of {\it Gaia} satellite data \citep{2005RvMA...18..179B,2005Sci...307.1440B,2008ApJ...678..187V,2012ApJ...753....7S,
 2019ApJ...872...24V,2021MNRAS.507.2592S}. With transverse motion data,  it is possible to reconstruct the past and future 
tri-dimensional trajectories in the sky with some uncertainties. 
The most recent orbital analysis, based on the transverse motion that the {\it Gaia} satellite has measured, casts  doubts on a close M31-M33 interaction in the past.
 A much higher probability for a past interaction has, however, been found using    
the M31 transverse motion resulting from a maximum likelihood method in connection with Local Group cosmological simulations and with estimates of  M31 satellite distances \citep{2016MNRAS.456.4432S}.
In Appendix A of  this study, we summarise, in detail, the results of several papers dedicated to Local Group encounters and emphasise the most relevant limitations and the need for  the present study.

It is well know that the fate of galaxies during encounters depends on  the masses of individual members of a gravitationally bound group. Galaxy's masses are therefore important ingredients for galaxy trajectories and pericentre times. Determining the dynamical mass of galaxies is possible only through accurate measurements of rotation curves and
satellite motion. For M31 and the MW, it has also been possible to estimate their total mass thanks to the numerous satellites 
\citep{2000MNRAS.316..929E,2000ApJ...540L...9E,2010A&A...511A..89C,2010MNRAS.406..264W,2011MNRAS.414.2446M,2014MNRAS.445.2049C}.
In previous orbital analysis, the M33 mass was often underestimated.
Luckily M33 has  a very extended HI disk which allows for the rotation curve to be traced well beyond the optical edge of the galaxy,
out to about three optical radii. The  analysis of the high resolution velocity field of the disk and of the stellar 
mass distribution via synthesis models has lowered the uncertainties on the dark matter distribution and baryonic mass of M33  \citep{2014A&A...572A..23C} . 

In this paper we examine, in detail, the past orbits of the brightest members of the Local Group, the MW, M31, and M33,
that are compatible with the most recent dark matter mass measurements
and with  Early Third Data Release of {\it Gaia} satellite (hereafter {\it Gaia}-EDR3)
 \citep{2021MNRAS.507.2592S}. We examine, in detail, the possible trajectories  that bring the M33
disk  in the actual configuration and see if the disk warp and the satellite distribution 
around M33  could have resulted from a close passage around M31 in the past. 
The paper is organised as follows. In Section 2 we summarise   the
proper motion results, galaxy dynamical masses,  and the M33 morphology in the outer regions.  In Section 3 we 
define the physical processes considered for orbit reconstruction and describe the type of numerical simulations used.
In Section~4 we discuss the orbital history of Local Group spiral galaxies.  In Section 5 we investigate if 
the outer disk distortion and  the non homogeneous distribution of gas clouds are tidally induced features of  a past M31-M33 encounter or signatures of gas accretion. We summarise our conclusions in Section~6. 
In  Appendix A we give a short summary of  previous analyses and results concerning the orbital evolution of M31 and M33, highlighting their limitations and the need 
for the study presented in this paper.  In Appendix B we describe, in detail, the N-body simulations used to find mass losses and the dynamical friction coefficient that were adopted in our semi-analytical approach.

\section{Observed properties: Mass, motion, and morphology of Local Group spirals}

In order to establish the past trajectories of the Local Group brightest members we need to know their masses, the three-dimensional
velocity vectors and  distances at the present time. We briefly summarise the most relevant mass   and distance estimates below. In the second part of this 
Section we analyse the proper motion of M31 and M33 and their radial velocities in selected reference systems. 
 We conclude this Section describing the most relevant morphological features of  M33 outer regions.

\subsection{Masses and distances}
 
Individual galaxy masses can be determined via rotation curve analysis or satellite motion but 
in the case of the Local Group there is another constraint that can help in determining the masses of the brightest members 
and that is the determination of the Local Group mass via timing argument or numerical simulations of structure formation in
a given cosmological context. 

The total  mass, that is often used in this paper, is defined 
as the sum of the baryonic and dark matter mass of a galaxy within its dark halo viral radius. 
The total mass of M31 and M33 has  recently been determined using dynamical analysis of their rotation curve traced
via high resolution and sensitivity 21-cm imaging of the atomic gas \citep{2010A&A...511A..89C,2014A&A...572A..23C}. Although
the rotation curve data only extend out to a maximum sampled radius, in these papers   more extended dark halo models have been tested 
that allow for the total virial mass of the galaxy to be recovered. 
The dynamical analysis of the M33 extended rotation curve shows that a dark matter halo, as predicted by
numerical simulation of structure formation in a hierarchical $\Lambda$CDM universe, is present and that
its  mass is non negligible compared to M31 and MW masses.
The lower boundary of the M33 mass listed in Table~1 is about 
twice the value of what can be inferred from the outermost sampled radius in the rotation curve data. 

The presence of bright satellite galaxies around the MW and M31 allows for
virial masses to be inferred using methods different than rotation curve analysis. 
Constraints on the masses of the largest Local Group galaxies can be drawn also by abundance matching (i.e. using results of numerical simulations in a given 
cosmological context)  joined by transverse motion measurements  
\citep{2014ApJ...793...91G,2016MNRAS.457..844F,2017MNRAS.465.4886C,2008MNRAS.384.1459L}. 
Table~1 summarises some of the most recent results concerning the mass determination of Local Group spirals.

\begin{table*}
\caption{Mass estimates of Local Group spirals.} 
\centering                                       
\begin{tabular}{c c c c }           
\hline\hline 
Object &  Mass &  Mass type and Method & References  \\                  
\hline\hline 
M33   & 0.3--0.5$\times 10^{12}$~M$_\odot$    &   Virial DM, Rotation curve & (1),(2)  \\
M31   & 1.0--1.6$\times  10^{12}$~M$_\odot$   &   Virial DM, Rotation curve & (3) \\
M31   & 1.0--1.8$\times 10^{12}$~M$_\odot$    &   Total M300kpc, Satellites     & (4) \\ 
M31   & 0.5--2.1$\times  10^{12}$~M$_\odot$   &   Virial DM, Satellites  & (5) \\
M31   & 1.6--2.6$\times  10^{12}$~M$_\odot$   &   Virial DM, Streams    & (6) \\
M31   & 1.0--1.7$\times  10^{12}$~M$_\odot$   &   Total M200 Globular clus. & (7) \\
M31   & 0.7--0.9$\times  10^{12}$~M$_\odot$   &   Virial DM, Escape velocity & (8) \\
M31   & 2.1--4.3$\times  10^{12}$~M$_\odot$   &   Virial DM, Satellites  & (9) \\
MW  & 1.1--1.7$\times 10^{12}$~M$_\odot$   & Virial DM Satellites  & (10) \\ 
MW  & 0.3--1.4$\times 10^{12}$~M$_\odot$   & DM200 Satellites   & (11) \\ 
MW  & 1.0--1.4$\times 10^{12}$~M$_\odot$   & Virial DM Distribution satellites & (12) \\
MW  & 1.0--1.5$\times 10^{12}$~M$_\odot$   & Virial DM Rotation curve & (13) \\ 
MW  & 0.8--1.0$\times 10^{12}$~M$_\odot$   & Virial DM Rotation curve & (14) \\ 
MW  & 0.7--1.0$\times 10^{12}$~M$_\odot$   & DM200 Rotation curve   & (15) \\
MW  & 0.8--1.2$\times 10^{12}$~M$_\odot$   & DM200 Rotation curve  & (16) \\
MW  & 0.4--0.7$\times 10^{12}$~M$_\odot$   & Virial DM Sag. Streams  & (17) \\
MW &  0.7--1.2$\times 10^{12}$~M$_\odot$   & DM200 Escape velocity & (18) \\
MW &  0.6--0.9$\times 10^{12}$~M$_\odot$   & DM200 Escape velocity & (19) \\
MW &  0.5--1.1$\times 10^{12}$~M$_\odot$   & Total M, Proper motion GC + RC & (20)\\
MW &  1.1--2.0$\times 10^{12}$~M$_\odot$   & Virial DM, Spherical Jeans Equation  & (21) \\
MW &  0.8--1.3$\times 10^{12}$~M$_\odot$   & DM200 Distribution halo stars   & (22)\\
MW+M31    & 2.6--3.7$\times  10^{12}$~M$_\odot$   & Timing + other arg. & (23) \\  
MW+M31    & 3.4--5.3$\times  10^{12}$~M$_\odot$   & Timing + other arg. & (24) \\  
MW+M31    & 1.6--3.6$\times  10^{12}$~M$_\odot$   & Matching + Velocities  & (25) \\  
MW+M31    & 1.4--3.4$\times  10^{12}$~M$_\odot$   & Matching + Velocities  & (26) \\  
MW+M31    & 1.3--4.4$\times  10^{12}$~M$_\odot$   & Matching + Velocities  & (27) \\  
MW+M31   &  2.8--6.9$\times  10^{12}$~M$_\odot$   & Likelihood free inference & (28) \\
\hline\hline 
\end{tabular}
\tablebib{
(1)\citet{2014A&A...572A..23C}; (2)\citet{2017AJ....154...41K}; (3)\citet{2010A&A...511A..89C}; (4)\citet{2010MNRAS.406..264W}; (5)\citet{2012ApJ...752...45T}; 
(6)\citet{2013MNRAS.434.2779F}; (7)\citet{2013ApJ...768L..33V}; (8)\citet{2018MNRAS.475.4043K};(9)\citet{2023ApJ...948..104P};
(10)\citet{2010MNRAS.406..264W}; (11)\citet{2014MNRAS.445.2049C}; (12)\citet{2020ApJ...894...10L}; (13)\citet{2011MNRAS.414.2446M};
(14)\citet{2016MNRAS.463.2623H}; (15)\citet{2020JCAP...05..033K}; (16)\citet{2020MNRAS.494.4291C}; (17)\citet{2014MNRAS.445.3788G};
(18)\citet{2019MNRAS.485.3514D}; (19)\citet{2022ApJ...926..189N}; (20)\citet{2022MNRAS.510.2242W}; (21)\citet{2020MNRAS.494.5178F}; 
(22)\citet{2021MNRAS.501.5964D}; (23)\citet{2012ApJ...753....8V}; (24)\citet{2023ApJ...942...18C};
(25)\citet{2016MNRAS.457..844F}; (26)\citet{2017MNRAS.465.4886C};
(27)\citet{2014ApJ...793...91G}; (28)\citet{2021PhRvD.103b3009L}.
}
\label{tabmass}
\end{table*}

\noindent
Most of the mass estimates given by the references in Table~\ref{tabmass} are relative to the dark matter mass of the halo within the viral radius. 
The mass within the viral radius,  defined as the radius inside which the mean density is about 100 times the critical density, is a factor 1.2 larger 
than M$_{200}$, the mass within a sphere with average density 200 times the critical density \citep{2001A&A...367...27W}. 
The baryonic mass (stars+gas) of M33 is negligible with respect to the dark matter mass while for the MW and M31 we should add to the dark matter mass
0.07 and 0.14 $\times 10^{12}$~M$_\odot$ baryonic matter respectively  in order to estimates the total mass
\citep{2010A&A...511A..89C,2011MNRAS.414.2446M,2012A&A...546A...4T,2016MNRAS.463.2623H,2014A&A...572A..23C,2020MNRAS.494.4291C}. 
In analysing the past motion of the three Local Group spirals  we considered several mass models. We first excluded the most extreme values for M31 and the MW mass listed in Table 1,  and considered   the
following  total mass ranges, that are consistent with several dark matter mass estimates at viral radius, plus baryonic mass estimates available in the literature: 

\begin{eqnarray}
 0.3\times 10^{12} \le &  M_{M33} & \le  0.5\times 10^{12} M_\odot, \\
 0.6\times 10^{12} \le &  M_{MW}  & \le  1.8\times 10^{12} M_\odot,\\
 0.8\times 10^{12} \le &  M_{M31} & \le  2.2\times 10^{12} M_\odot   
 \label{st}  
.\end{eqnarray}

We sampled total masses randomly, considering a flat distribution over the above intervals or assuming Gaussian distributions centred at 0.4, 1.2, and 1.5$\times 10^{12}$~M$_\odot$  truncated at $3\sigma$, with a dispersion 
$\sigma$ of 0.03, 0.20 and $0.23\times 10^{12}$~M$_\odot$ for M33, MW and M31 respectively.  Between all possible combinations we always considered cases for which the MW mass never exceeds the M31 mass  (i.e. we disregarded random mass triplets that did not satisfy this condition). 

Using N-body simulations we inferred that a close passage of M33 around M31 changes the matter distribution in these two galaxies. This change is because some of the matter bound to M33 leaves the galaxy after the encounter and is accreted by M31 or it is lost in space.
Variations of the viral mass are negligible for M31, but for M33 the mass decreases by 35-40$\%$. To account for this loss, 
we assumed that the M33 mass increases during a backward orbit integration according to the following law:

\begin{equation}
M_{M33}(t)=M_{M33}^0\times(2-exp(-t/t_{max}))
\label{m33loss}
,\end{equation}

where M$_{M33}^0$ is the M33 mass today and t$_{max}$ is the interval of time to reach the initial conditions for the forward integration. This additional mass is lost during the forward integration.
We consider an interval  t$_{max}$=9.2~Gyr in computing the position and velocity of the galaxy centre of mass back in time. 
This interval of time is chosen because we would like to integrate back in time as much as possible to trace past encounters, and at the same time have the galaxies already formed according to the hierarchical cold dark matter formation scenario. In term of galaxy mass assembly we selected the redshift for which the specific halo mass increase rate ($\Delta$M/$\Delta$t, normalised by the halo mass at a given specific cosmological time) is small. According to \citet{2023ApJ...959....5L}  (see their Figure 6) the specific mass  increase rate becomes less than 0.2 Gyr$^{-1}$ approximately 9.2 Gyr ago, that is  galaxies at this cosmic time have masses close to today's masses, and regular disk galaxies are already  a dominant population \citep[e.g.][and references therein]{2019MNRAS.487.1795S}.   

If M33 had a pericentre passage around M31 in the past, we increased the M33 mass during  the backward integration by about 63$\%$ in a time t=t$_{max}$=9.2~Gyr. This increase accounts for the total mass loss experienced by M33 during the close encounter in the forward integration, as shown in detail in Appendix B with the use of N-body simulations.
Mass loss along an orbit is not a continuous function of time but it happens soon after the pericentre passage (see Appendix B). We cannot  improve the mass-loss rate approximation not knowing apriori the pericentre time. Time integration along orbital trajectories considering mass losses for  M33, with the M31 and MW masses constant in time and for masses in the ranges given by Eq.(1), (2), and (3) at t=0, is considered the {standard} mass model.

We also tested  some alternative mass models. The first one is  a {low-mass} model for M33  that
predicts that the mass of M33 today is between 1.8 and 3$\times 10^{11}$~M$_\odot$. In other words the mass derived through dynamical analysis is interpreted as the original mass 9.2 Gyr ago, because the close passage around M31 removes  mass from  the outer halo, beyond the region tested by the rotation curve analysis. For this reason {low-mass} mass values are below those listed in Table~1, inferred from extrapolations of dynamical models of the rotation curve. For the {low-mass} model   the M31 and MW masses are constant through time while the M33 mass increases during the backward integration and decreases from 9.2~Gyr ago to the present time according to Eq.(\ref{m33loss}). The {standard} and {low-mass} models have been used for the centre of mass back orbit integration as well as for test-particle simulations.

For the analysis of the centre of mass orbits we also considered  the possibility that the M33 mass is constant with time, because mass gained by small accretion events roughly balances mass losses due to the interaction with M31 in the past. We coupled this mass model for M33 to a time varying mass model for the MW and M31.  Galaxy  growth with time and the MW and  M31 have likely accreted mass in the time interval we are considering. This is particularly relevant for M31 since this galaxy is disturbed due to recent merger events:  in particular both observations and cosmological simulations predict that M31 has undergone a major accretion event less than 9 Gyr ago \citep{2018NatAs...2..737D,2022MNRAS.516.5404S}. 
In the back orbit integration we used the following equation to account for the possible decrease in mass of MW and M31 back in time:

\begin{equation}
M_{MW,M31}(t)=M_{MW,M31}^0\times(1-f_{MW,M31}+f_{MW,M31}\times exp(-t/t_{max})),
\label{mloss}
\end{equation}

where M$_{MW,M31}^0$ indicates the MW or M31 mass today and f$_{MW,M31}$ is a constant used to model the mass evolution of the MW or M31 through time. We find interesting to use this mass model for investigating the past orbits of Local Group spirals if M31 today is more massive than the range given by Eq.(\ref{st}), as some recent study suggested \citep{2023ApJ...948..104P}. In particular we refer to the  highvar mass model  when  $1.8\times 10^{12} \le   M_{M31}  \le  3.2\times 10^{12}$~M$_\odot $ with $f_{M31}=0.75$  and $f_{MW}=0.5$, with the M33 and MW mass constant through time  and within the range given by Eq.(1) and (2). For this model  M31 has a similar or  higher mass than the MW  through time, and the MW has a mass of the same order or higher than the M33 mass.  We used this mass model mostly to investigate possible variations of the pericentre probability. This model has not been used in N-body computations or tested through cosmological simulations.

Given the uncertainties on the M31 and M33 distances \citep[e.g.][and references in the NASA/IPAC Extragalactic Database]{2021ApJ...920...84L,2021MNRAS.508.3035S} we considered also a 
Gaussian or random distribution for M31 and M33 distances with dispersion of 15 and 20~kpc and centre values of 783 and 840~kpc respectively  in close agreement with mean values 
of distance modulus measurements listed in the NASA/IPAC Extragalactic Database and with  \citet{2001ApJ...553...47F}. 
  
\subsection{Three-dimensional velocities of M33 and M31}

The motion of M33 and M31 with respect to the barycentre of the Local Group or to the centre of the MW can be measured
only through an accurate determination of the radial and tangential velocities with respect to the Sun and  of the Sun motion with
respect to the MW centre. Tangential velocity, or proper motion,  is the change 
in angular position of a compact object in the sky as seen from the Sun and it can be measured today
only for objects in the Local Group. For M33 and M31 these velocities have been measured  by detecting the 
proper motion of water masers, or of satellites and stars  in HST fields. Recently, more accurate measures have been possible thanks to 
{\it Gaia} satellite data for bright stars in these galaxies. These are especially relevant for M31 for which the absence of strong water masers limits the use of these to infer proper motion.
To determine the velocity of the galaxy hosting the objects one has to determine the internal motion of the objects with respect to
the galaxy centre and the distance and motion of the Sun with respect to the MW centre.
The large measurement uncertainties and the various corrections needed to derive the final tangential velocities 
implies that we are left with a wide range of values. We summarise in Table~\ref{tabpm} the observed components of the proper motion 
of M33 and M31 with their uncertainties in the heliocentric rest frame. We define a positive proper motion in the western direction 
when an object is moving towards the west and hence it has the opposite sign of RA shift. We list the centre of mass  values
for each velocity component, $\mu_W^{com}$ and $\mu_N^{com}$,  obtained after inner disk velocities, such as the
rotational motion, have been subtracted. For M33 the proper motion measurements are from observations of water masers \citep{2005Sci...307.1440B}. We took the average
velocity of the two masers after we corrected  for the disk rotation. We used the rotation curve of \citet{2014A&A...572A..23C}, that gives slightly different values than
those quoted by \citet{2005Sci...307.1440B}, and which
implies V$_{rot}$=108~km~s$^{-1}$ at a the radial distance of 5.3~kpc for IC133 and V$_{rot}$=93~km~s$^{-1}$ 
at a radial distance of  2.6~kpc for M33/19. The components of the rotational velocities  are $\mu_W^{rot}$(IC133)=16.36~$\mu$as~yr$^{-1}$
$\mu_N^{rot}$(IC133)=-20.21~$\mu$as~yr$^{-1}$ and $\mu_W^{rot}$(M33/19)=-10.02~$\mu$as~yr$^{-1}$
$\mu_N^{rot}$(M33/19)=-10.57~$\mu$as~yr$^{-1}$.   Throughout this paper we used the weighted mean shown in the last entry of Table~2.
For M31 star fields we report the results of the weighted average of the three fields examined by \citet{2012ApJ...753....7S} with the relative
 corrections for internal kinematics as given by  \citet{2012ApJ...753....8V}. For M31 satellite kinematics we applied
no corrections to the values reported in \citet{2012ApJ...753....8V}. From the {\it Gaia}-EDR3 data analysed by \citet{2021MNRAS.507.2592S}, we 
selected the velocities traced by the sample of blue stars that suffer less contamination by local sources.  We used these transverse velocity components throughout this study. As Table~2 shows, these values are very similar to measures based on HST data.

\begin{table}
\caption{Proper motion in the sky plane.} 
\centering                                       
\begin{tabular}{c c c c c }           
\hline\hline 
Object & Method &  $\mu^{com}_W$ &  $\mu^{com}_N$  & Ref.\\  
  &  &$\mu$as~yr$^{-1}$&$\mu$as~yr$^{-1}$ & \\                
\hline\hline 
M31-a  & Satellite Kinematics          &  -27$\pm$11 & -12$\pm$10  & (1) \\
M31-b   & HST star fields                  & -45$\pm$13 & -32$\pm$12  & (2) \\
M31-c   & Weighted mean a+b    & -34$\pm$9  & -20$\pm$8  & (2) \\
M31-d   & Stars {\it Gaia} DR2               & -65$\pm$24 & -57$\pm$22 & (3) \\
M31-e   & Weighted mean  b+d   & -49$\pm$11 & -38$\pm$11 & (3) \\
M31-f  & Stars {\it Gaia} EDR3 blue        & -49$\pm$11 & -37$\pm$8 & (4) \\
M33-g   & Water Masers IC133      & -21$\pm$8   & 6$\pm$9   & (5,6) \\
M33-h   & Water Masers M33/19     & -25$\pm$8   & -2$\pm$9  & (5,6) \\
M33-i  & Weighted mean h+g    & -23$\pm$6   & 2$\pm$6 &  (5,6) \\
M33-j   & Stars {\it Gaia} DR2              & -31$\pm$25 & -29$\pm$17 & (3) \\
M33-k   & Weighted mean i+j   & -23$\pm$6 & -1$\pm$6 &  (5,6,3)\\
\hline\hline 
\end{tabular}
\tablebib{
(1) \citet{2008ApJ...678..187V}; (2 )\citet{2012ApJ...753....7S}; (3) \citet{2019ApJ...872...24V};
(4) \citet{2021MNRAS.507.2592S}; (5) \citet{2005Sci...307.1440B}; (6) \citet{2014A&A...572A..23C}.
}
\label{tabpm}
\end{table}

We  transformed the proper motion components relative to a reference system centred on the Sun, in the Galactocentric reference system. 
This system has the x-axis oriented from the Sun to the MW centre, with the Sun
on the negative x-axis, and the z-axis perpendicular to the galactic disk plane. For changing reference system we need to
know the Sun's peculiar motion and its  rotation velocity around the MW centre. The Sun's galactocentric distance is assumed to be 8.28$\pm0.04$~kpc \citep{2021A&A...647A..59G},
and the Sun's circular velocity V$_r\odot$=239$\pm$5~km~s$^{-1}$ \citep{2011MNRAS.414.2446M}.
For the peculiar Sun's velocity with respect to the Local Standard of Rest we adopted the estimates of \citet{2010MNRAS.403.1829S} 
that gives the velocity of the Sun pointing to the galactic centre U$_{pec}$=11.1$\pm1.2$~km~s$^{-1}$, in the direction of rotation 
V$_{pec}$=12.24$\pm2.1$~km~s$^{-1}$ and vertically out of the plane W$_{pec}$=7.25$\pm0.6$~km~s$^{-1}$. These values are
also in agreeement with other more recent measurements  \citep{2019ApJ...885..131R}.
The Sun's motion causes an apparent motion of M31 and M33, which is the reflex of the Sun's motion. 
The solar reflex motion, $\mu^\odot_W$=-39$\mu$as~yr$^{-1}$  $\mu^\odot_N$=-22$\mu$as~yr$^{-1}$, must be subtracted from the M31 and M33 RA and DEC 
centre of mass displacements. This correction implies that the values listed in Table~\ref{tabpm}, predict a nearly 
radial orbit of M31 towards the MW. Table~\ref{tabpm} shows that M31's proper motion measurements from different datasets are consistent; for this reason
we disregarded the significantly higher 
values of the M31's transverse velocity derived by  matching   M31 and its satellite kinematics with analogues in  $\Lambda$CDM 
cosmological simulation \citep{2016MNRAS.456.4432S} ($\mu^{com}_W=$ -9$\pm$19  $\mu^{com}_N$=5$\pm$16).

Radial velocity measurements in the optical and 21-cm give consistent results: -179$\pm$3~km~s$^{-1}$ for M33 and 
-308$\pm$8~km~s$^{-1}$ for M31. For the geometrical parameters of the M31 and M33 disks we assumed a position angle of 35$^\circ$ and 
inclination 71$^\circ$  fo M31 and  of 22$^\circ$ and 54$^\circ$ for M33. More details on the M33 structure are given in the
next subsection.

Proper motion angular velocities in the plane of the sky have  then been converted into  linear velocities. 
The velocities  in the Galactocentric reference system are labelled V$_{x,y,z}^{\odot}$; we then
subtract the Sun's motion to obtain V$_{x,y,z}$. The coordinates of M33 and M31 in the Galactocentric frame in kpc have
been derived after subtracting the Sun's distance from the galactic centre.
Standard deviations for the position of the centre of mass reflect  uncertainties in the distance measurement.

\subsection{M33 morphology and galaxy outskirts}

 The low-luminosity flocculent spiral galaxy M33 is the third largest member of the Local Group. It has  two weak  arms, not symmetrically winded, 
 and an interstellar medium rich of gaseous filaments extending out to about 7~kpc, to the edge of the star forming disk. Although the inner disk is relatively
undisturbed, the northern arm is less regularly shaped than the southern arm.  M33, is a bulgless galaxy with only two known optically bright dwarf galaxies nearby 
that are candidates for being its satellites: 
AndiXXII  \citep{2009Natur.461...66M,2016ApJ...833..167M} and Pisces VII \citep{2022MNRAS.509...16M}. Given its mass,
$\Lambda$CDM cosmological simulations predicts that M33 should host a much larger number of satellites, at least 10 with baryonic mass larger than 10$^3$~M$_\odot$.
 The HI disk is three times more extended than  the star forming disk and it is clearly
warped, as described in detail by \citet{1997ApJ...479..244C,2014A&A...572A..23C}. The outer disk has the
same inclination of the inner one with respect to our line of sight but the position angle of the major axis
changes by about 30 degrees from the inner disk and it is more alligned with the M31 direction.
While the undisturbed inner disk of M33 indicates that no major collisions between M31 and M33 or between M33 and a satellite
has taken place in the past, the warp might be the result of a close pass by. An alternative possibility is the gas accretion process, 
described in Section~1 and discussed in Section~5. The gaseous mass of M33 is about half of its stellar mass. The last one has been inferred via population synthesis
models to be about 4.8$\times 10^9$~M$_\odot$ \citet{2014A&A...572A..23C}. A deep optical analysis has shown that a stellar component is associated with the outer
disk, corresponding to a star forming episode about 100-300~Myrs ago \citep{2011A&A...533A..91G}.  The PAndAS 
survey  of M31 and its environment shows also that the M33 outer disk has a stellar component with stars of
different ages, some older than 1~Gyr \citep{2010ApJ...723.1038M}.

Atomic gas has been observed in large cloud complexes and small isolated clouds in the circumgalactic medium of M31 and M33
\citep{2004A&A...417..421B,2004ApJ...601L..39T,2008A&A...487..161G}. Gas clouds can result from gas accretion or  can be the baryonic component of dark satellites, which have no  stellar counterparts  \citep{2008A&A...487..161G}. No stellar structure correlates with HI detections in the M33 environs   \citep{2009Natur.461...66M,2011A&A...533A..91G}. A few  clouds appear to 
to be located in proximity of the giant stellar stream of metal-rich stars between M31-M33 
\citep{2007ApJ...671.1591I,2012AJ....144...52L,2013ApJ...763....4L}, but many other around M33 are on the opposite 
side with respect to M31 and rather massive. If these gaseous clouds are the baryonic counterparts of a population of dark 
satellites, whose gas has been preserved in the dark matter halos but has never condensed to form stars, 
one should explain their non isotropic distribution around M33.

\section{Equations of motion and numerical simulations}

We simulated the encounter between M33 and M31   after using a backward integration scheme to recover the initial conditions 9.2~Gyr ago. 
Both M33 and M31 are bound to the Local Group and we included in the simulations the MW    to be able to reproduce the observed positions and  velocities of M31 and M33, and the geometrical projection of the M33 disk on the plane of the sky at the end of the forward integration.

The backward integration has been carried out considering  the centre of mass of M31, of M33 and of the MW.
The baryons of M33 are distributed  all in the disk and we investigated its dynamical evolution using test-particles from 9.2~Gyr ago to the present time. These particles are collisionless and massless and feel the gravitational attraction of the M33 mass inside their orbits, of the MW and of M31.  
Although the use of an N-body+SPH numerical simulations would allow a more complete study of the M33-M31 system during their interaction, having these galaxies both gas and stars, we specify in what follows why we  rely on  test-particle and N-body simulations  for understanding if  the M33 warp formation and the non homogeneous spatial distribution of its dark satellites have been triggered by a past close encounter with M31.
 
If the warp represent the response of a galaxy disk to a recent close encounter, it is shaped by gravity and dissipative gas-dynamical processes cannot prevent or enhance it. This warp formation scenario can be examined considering the gravitational field of the baryons and dark matter halos. The use of N-body simulations of collisionless particles can complement such an approach to estimate analytical approximations such as the dynamical friction term and mass losses. 
The damping time of a warp can however be different between gas and dissipationless matter. Nevertheless we note the following: $(i)$ there are several indications that  collisional dissipation in a gas layer become relevant only when the differential precession is high \citep{2006MNRAS.370....2S}. Differential precession for warps is actually estimated to be small \citep{2020NatAs...4..590P}. In the unlikely case of a high precession rate, the gas will  quickly be driven to the original plane thus casting doubts on  positive conclusions about possible warp formation that, we will see, are never reached throughout this study. $(ii)$ Gas often condenses into compact clouds that behave quasi ballistically in a similar manner to stellar system, with the cloud-cloud collision time comparable with or greater than the orbital time \citep{1992ARA&A..30...51B}. This consideration is particularly relevant for M33 whose warp is made of gas and young stars, the last ones being born out of compact gas condensations  \citep{2011A&A...533A..91G}. $(iii)$ Collisionless and test-particle simulations are less time consuming and allow one to investigate the full parameter space of initial conditions compatible with the data.

In our computation we shall use two reference systems. One is the Galactocentric
orthogonal reference system with the origin at today's MW centre, the x-y plane along the galactic plane and
the Sun along the negative x-axis. However, for visualising the results we also use the '{tangent}' reference system:
this has the origin at the M33 centre in the plane of the sky, the x-z plane tangent to the celestial sphere, with the x-axis along the
decreasing right ascension direction and the y-axis pointing radially away from the Sun.

\subsection{Centre of mass orbits and initial conditions}

For retrieving the possible initial conditions of  test-particle simulations, 9.2~Gyr ago, we integrated back in time the centre of mass orbits in the galactic orthogonal reference system, starting from today's positions and velocities of M31 and M33 described in the previous Section.  
These orbital initial conditions have been recovered  through a semi-analytic  integration (or back orbit numerical scheme) described below.

The dynamics of each galaxy in today's Galactocentric reference system is governed by the usual coupled equation of motions 
with the accelerations   {\bf a}$_i$  (with $i=MW,M31,M33$) given by:

\begin{equation}
\begin{aligned}
{\bf a}_{MW}&= -{\bf \nabla}(\phi_{M31}+\phi_{M33})\\
                        &=\sum_{i=M31,M33}  {GM_{i,MW}\over r^3_{i,MW}} ({\bf r}_i-{\bf r}_{MW}),\\
{\bf a}_{M31}&= -{\bf \nabla}(\phi_{MW}+\phi_{M33})-{{\bf f}_{DF}^{33}\over M_{M31}}\\
                       &= \sum_{i=MW,M33}  {GM_{i,M31}\over r^3_{i,M31}} ({\bf r}_i-{\bf r}_{M31})-{{\bf f}_{DF}^{M33}\over M_{M31}},\\
{\bf a}_{M33}&= -{\bf \nabla}(\phi_{MW}+\phi_{M31})-{{\bf f}_{DF}^{31}\over M_{M33}}\\
                       &= \sum_{i=MW,M31}  {GM_{i,M33}\over r^3_{i,M33}} ({\bf r}_i-{\bf r}_{M33})-{{\bf f}_{DF}^{M31}\over M_{M33}},
\end{aligned}
\end{equation}

\noindent
where G is the gravitational constant, {\bf r}$_i$ is the distance between the $i$-galaxy  and today's MW centre, and r$_{i,j}$ is the relative distance 
between the galaxy $i$ and the galaxy $j$ at a given time. The symbol M$_{i,j}$ is the mass for the galaxy $i$ at a distance  
r$_{i,j}$  from the galaxy $j$. The MW is approximated as a point source, since both M33 and M31 are at several hundreds of kpc  distance
 from the MW through the cosmic time we considered. 
A Navarro Frenk and White (NFW) dark matter profile \citep{1997ApJ...490..493N}  with  concentration of 12 and 9.5 has been used to fit the rotation curve of M31 and M33 respectively  \citep{2010A&A...511A..89C,2014A&A...572A..23C}.
However, we need to tune the semi-analytical computation with the N-body simulation and for the last one we prefer to use a Hernquist dark matter profiles \citep{1990ApJ...356..359H} that has a finite mass  (see Appendix B for more details on the N-body simulation). For this reason we carried out the semi-analytic orbit integration considering  Hernquist halo profiles that have the same enclosed mass, within the viral radius, as the NFW profile fitted to rotation curves. We derived the Hernquist scale radius $a_s$ according to the prescription summarised in \citet{2012ApJ...753....8V}.

The M31 dark matter halo is more massive and more extended than the M33 halo at all times but M33 mass is less than a factor ten smaller than M31 mass. Therefore we  considered that M33 experiences dynamical friction due to  M31 halo  but also  vice versa.  The friction exerted by M31 on M33 is {\bf f}$_{DF}^{M31}$, and that exerted by M33 on M31 is {\bf f}$_{DF}^{M33}$.
The dynamical friction  acts as a positive acceleration when we integrate back in time for a total of 9.2~Gyr to recover initial conditions. 
We approximated its value using the Chandrasekhar  formula \citep{1943ApJ....97..255C}

\begin{equation}
\frac{{\bf f}_{DF}} {M_i}  =
 \frac{4\pi G^2 M_i \rho(d) {\hbox{ln}} \Lambda }  {v^2}  
\biggl\lbrack  {\hbox{erf}}(X)-\frac{2X}  {\sqrt{\pi}}  {\hbox{exp}}(-X^2) \biggr\rbrack \frac {\bf v}{v}
\end{equation}
 
\noindent
where   M$_i$ is the mass of the galaxy experiencing the dynamical friction, $X=v/(\sqrt{2} \sigma)$ is the ratio between the relative velocity of the two galaxies  and the one-dimensional velocity dispersion at distance $d$ from its centre;  $\rho(d)$ is the halo density of the 
host galaxy. We evaluated $\sigma$ of the dark matter halo  following the approximation of \citet{2003ApJ...598...49Z}, relative to  a NFW dark matter halo, but also  using the analytical expression for an Hernquist halo \citep{1990ApJ...356..359H}. We don't find appreciable variations in the orbital solutions using both prescriptions, because of the similarity between the radial dispersion velocity profiles of the two halo models \citep[e.g.][]{2018MNRAS.476.2086L}, given also the large pericentre distances found for our initial conditions.

The Coulomb factor, ln$\Lambda$, is unknown and several parametrisation
exist. Unfortunately we cannot use the prescription  for Local Group galaxies encounters
given by \citet{2012ApJ...753....9V} since the authors considered only cases of galaxies with equal masses or with a mass ratio of 1:10 such as a small satellite orbiting a more massive galaxy. 
In our case the M33 mass,  is smaller than the M31 mass but it is larger than 0.1 M$_{M31}$. To make the semi-analytical computation in agreement with the N-body computation 
we use  a similar parametrisation to what \citet{2012ApJ...753....9V}  used for the unequal mass case, namely

\begin{equation}
{\hbox{ln}}\Lambda=max\biggl[0,{\hbox{ln}} \biggl(\frac{C\  r_{31,33}} {a_s}\biggr)\biggr],
\end{equation}

\noindent
with C a numerical constant, determined through a comparison with results of numerical N-body simulations, and a$_s$ the M33 Hernquist halo profile scale length. 
The N-body simulation used to determine C is described in detail in Appendix~B and its use can be summarised as follows.  We guess a value of C and recover for this value the initial conditions 9.2~Gyr ago using semi-analytical computation (back orbit centre of mass integration). We then use these conditions for the forward integration using both a test-particle simulation and an N-body simulation. We compare the two solutions and vary C accordingly repeating  the procedure until the orbital solutions of the two numerical approaches converge. As shown in Appendix-B the value of C is not universal  and varies by varying the initial conditions of the orbits. However,  we have found C in a very narrow range, 0.85--1.05, and we adopt the value C=1.
 
The  value of C depends slightly on whether the dynamical friction of the M33 halo on the M31 orbit is considered or neglected.  If one considers the dynamical friction of both M31 and M33 halos then the value C=1 gives a good match between the semi analytical orbit computation and the N-body simulation. The value C=0.82 found by \citet{2012ApJ...753....9V} seems more suitable for orbital solutions when {\bf f}$_{DF}^{M33}=0$. We underline that halo virial radii are large, of the order of 300 and 190~kpc for M31 and M33 respectively, so halo particles of M31 and M33 overlap  and dynamic friction cannot be neglected.
 
The time integration has been  carried out using leapfrog scheme. This is fully time reversible and thus appropriate to ensure that the backward and forward integration of the orbits are consistent with one another.  Typical time step are of the order of 4$\times 10^{-4}$~Gyr.
We define the pericentre of the M33 orbit as the minimum distance between M33 and M31, D$_{p}$, reached at a time t$_{p}$. 
Knowing the centre of mass orbits we can determine (a) the number of pericentre passages (b)the time at pericentre passage (c) the pericentre distance, using the semi analytical back orbit integration scheme.

\subsection{Test-particle simulations}

Test-particle simulations have been designed to study the impact of a close pericentre passages of M33 around M31 on the M33 disk, using a variety of initial conditions recovered through back orbit integration of the M31, MW and M33 centre of mass. For test-particle simulations we refer to the conditions at the beginning of the forwards time integration, 9.2~Gyr ago, as initial conditions. We instead refer to t=0, or present time conditions, as the final results. We adopted 10000 particles to simulate the M33 disk, that extends for 20~kpc in radius and it is  at the centre of the M33 dark matter halo. At the beginning of the simulation all test-particle follow  coplanar orbits in a disk  oriented spatially  as  today's  bright star forming disk of M33, as  seen in the tangential frame. The centre of mass of M33, MW and M31 have initial positions and velocities equal to that recovered through back orbit integration. 
The motion of the galaxies in the Galactocentric reference system  are described by Eq.~(6), and are coupled to the following additional equation of motion for each $k$-test-particle:    

\begin{equation}
{\bf a}_{k}= -{\bf \nabla}(\phi_{MW}+\phi_{M31}+\phi_{M33})=\sum_{i=MW,M31,M33}  {GM_{i}\over r^3_{i,k}} ({\bf r}_i-{\bf r}_{k})  
\end{equation}

\noindent
where $r_{i,k}$ is the distance between the $k$-particle in M33 disk and the centre of mass of the $i$-galaxy ($i$=MW, M31 or M33). Galaxy masses  are total masses (dark matter+baryonic mass) inside a radius $r_{i,k}$. For our initial conditions test-particles are never so close to the MW and M31 centre of mass as to intersect the disks of these galaxies where most of the baryons are. For  M33 instead this clearly happens, therefore we considered both the mass of the dark matter halo and the mass of the baryonic disk only up to a radial distance $r_{33,k}$. The baryonic mass  inside $r_{33,k}$ has been computed using Figure~10 of \citet{2014A&A...572A..23C}, assuming spherical geometry for its gravitational potential. This  approximation is justified by the large radial distance where the warp settles, beyond six disk scale lengths. In this region the surface density of the baryons is low and dark matter potential dominates.  

The use of N-body computations to simulate the past  encounter between  M31 and M33 has shown that the dark matter halo of M33 looses mass after a pericentre passage. Mass losses are from the outer regions of the M33 halo but the halo keeps its particles and  its original density distribution in the inner regions. In the test-particle simulations we have therefore decreased the halo mass with time according to Eq.~(\ref{m33loss}). At each time step we recomputed the virial radius  and scale length of the Hernquist halo model equivalent to an NFW halo, keeping the concentration parameter fixed.

We considered  all possible orbits consistent with today's data described in Section~2.
At the end of each test-particle simulation we compared the  particle spatial distribution and velocities, in the tangential coordinate system, 
with the observed baryonic distribution on the plane of the sky and the observed line of sight 2D-velocity map of M33. 
We searched for orbits that are able to generate and maintain a warp in the M33 disk 
similar to what is observed today (tilt of PA clockwise by 30-40~deg and almost no tilt in the inclination).

Similarly, we used  test-particle simulations to investigate  the spatial distribution an ensemble of
satellites orbiting M33, while this galaxy moves along the orbit. Our goal is to check if satellites are removed or their distribution skewed in any particular direction due to a past encounter with M31. 
We used one test-particle for each satellite,  being these of negligible mass compared to M33 mass,   and distributed the satellites  isotropically  around M33 at a fixed radial distance, that we considered between 40 and 70~kpc.

\section{The orbital history and the probability of a past pericentre passage}

We integrated the orbits of the three Local Group spirals back in time, for the last 9.2~Gyr. There are two relevant orbital solutions. For the first one  the
minimum distance between M33 and M31 is now, that is to say the two galaxies are now approaching for the first time and their separation will decrease further in the near future. The second one for which M33 has been closer to M31 in the past: in this case M33 did go around M31 experiencing a pericentre passage during the last 9.2~Gyr,  the distance between the two galaxies decreased, reaching a minimum value D$_{p}$ at t$_p$ and then increased again for a subsequent interval of time.

Given the most likely values of  physical parameters listed in the previous sections, the probability of a pericentre passage of M33 around M31 in the last 9.2~Gyr is between 11$\%$ and 27$\%$.
Variations depend on the mass intervals considered and on whether  these masses are randomly sampled in the given ranges or they follow Gaussian distributions
around the mean. For the standard and low-mass mass model a higher probability is found for randomly sampled masses.
The lowest probability is instead relative to Gaussian distributions for the {low-mass} M33 model. Table~3 lists the pericentre passage probability, 
the mean of the pericentre distance and pericentre time, and their standard deviations, according to Gaussian and random sampling for some of the mass models considered. We can see that the mean pericentre distance and time and their dispersion  are rather similar independently 
of the mass model and mass distribution in the range considered. Considering  a large mass for M31, as in the highvar mass model,  increases considerably the probability of a past pericentre.

Distance variations do not  affect  the probability of having a close passage in the past. For a Gaussian distribution of  the M33 distances, around a mean value of 840~kpc with a dispersion of 20~kpc,  the mean pericentre separation increases to 154~kpc  for M33 distances larger than the mean, or decreases to 123~kpc for M33 distances smaller than the mean. 
Smaller variations, but in the opposite direction, are found by varying the M31 distance. We show in Figure~1 the distribution of pericentre distances for the more general case that includes M31 and M33 distance dispersion around the mean values and for the case of distances fixed at their mean value  (100000 cases examined). Figure~1 shows a correlation between the pericentre distance and pericentre time when the passage happens earlier than 5~Gyr ago. 

\begin{table}
\caption{Probability of a past pericentre P$_p$ of M33 around M31 and mean values of D$_p$ and t$_p$ considering f$^{M33}_{DF}\ne 0$, f$^{M31}_{DF}\ne 0$, and  C=1.0. Today's distances of M31 and M33  have been held fixed to their mean value.} 
\centering                                       
\begin{tabular}{c c c c}           
\hline\hline 
Sampling and &   P$_{p}$ &  D$_{p}$&  t$_{p}$  \\  
  mass model &  & kpc & Gyr  \\                
\hline\hline 
Gaussian-standard      &  0.18 &  129$\pm$18  &   6.2 $\pm$ 2.0 \\
Gaussian-low-mass   &  0.11 &  129$\pm$16  &   6.3$\pm$ 2.0 \\
Random -standard      &  0.27 &  128$\pm$19  &   6.0$\pm$ 1.9 \\
Gaussian-highvar   &  0.46 &  122$\pm$21  &   5.6$\pm$ 2.0 \\
\hline\hline 
\end{tabular}
\label{tabperi}
\end{table}

\begin{figure} 
\includegraphics [width=0.55\textwidth, height=0.65\textwidth]{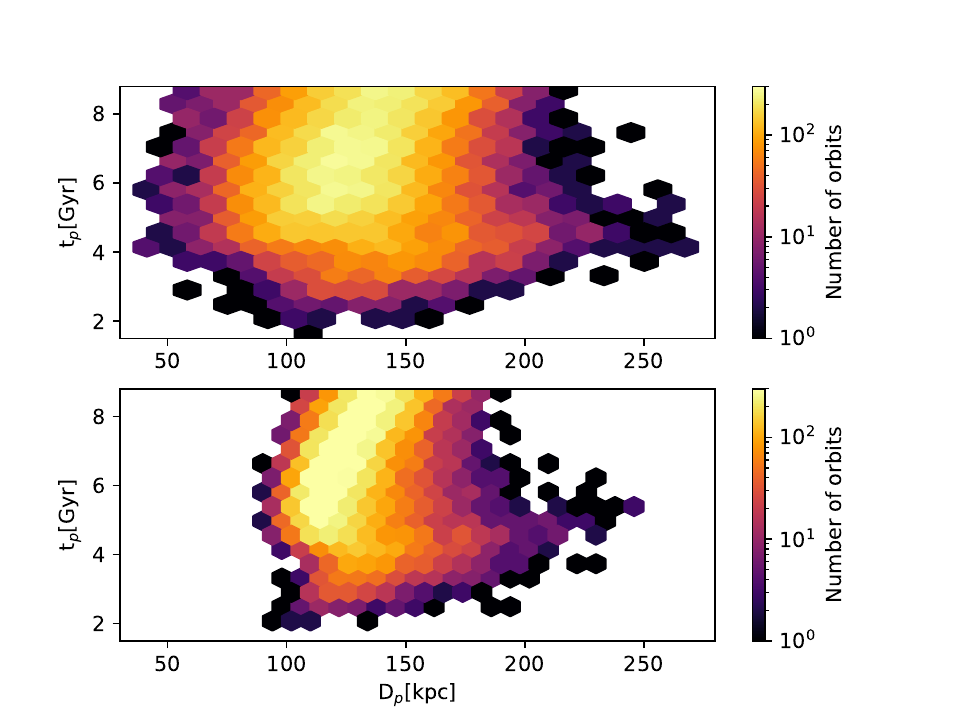}
\caption{ Distribution of pericentre distances D$_p$ and times t$_p$ for M33 orbits along which M33 had a close passage around M31. These are shown for today's distances to M31 and M33 equal to their mean value in the bottom panel while in the top panel  distance dispersions around the mean are considered. Regions in the D$_p$-t$_p$ plane are colour coded according to the number of orbital solutions with a close pericentre passage in the past. A total of 100000 possible orbits have been examined.
}
\label{dperi}
\end{figure}

\begin{figure*} 
\hspace{-0.7cm}
\includegraphics [width=1.0\textwidth]{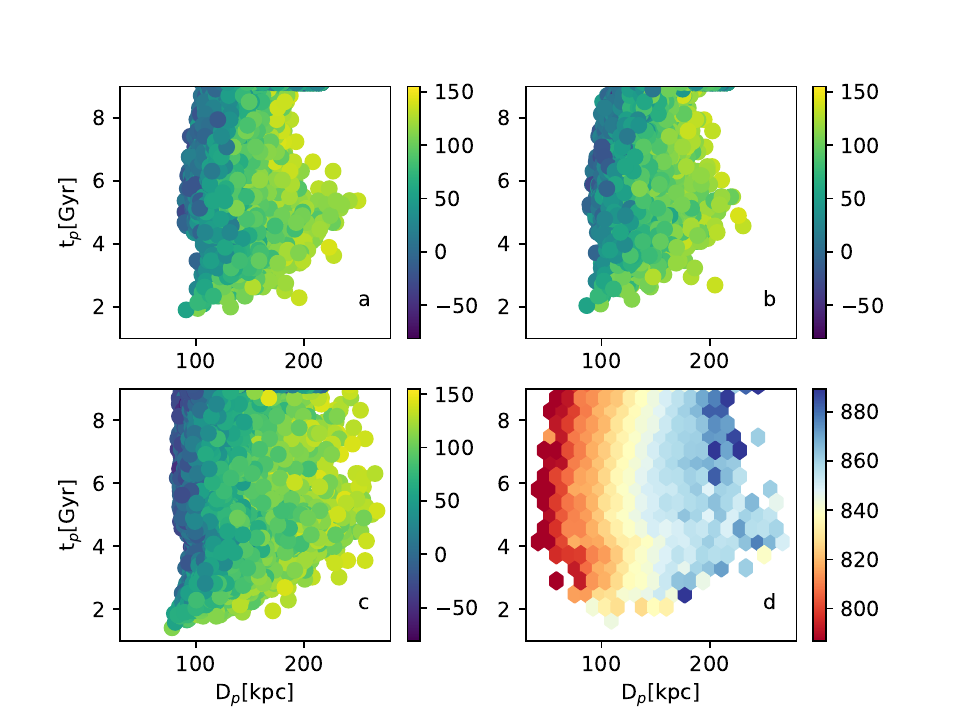}
\caption{Pericentre times in gigayears versus pericentre distances in kiloparsecs for M33 close passages around M31. Time increases from today to early cosmic times. In panel $a$ we show the standard case for M31 and M33 mass distributions, with M33 mass larger in the past. In panel $b$ we consider the low-mass model,  in panel $c$  the highvar mass model.   In panels $a,b,c$  the most likely values for today's distances of M31 and M33 from the MW are considered. Points are colour coded according to today's M31 velocity along the x-axes in the Galactocentric frame,  V$_{x}^{M31}$,   shown by  the colour bar in units of km~s$^{-1}$. In panel $d$ we consider variations of M31 and M33 distances around their mean values and  the standard model for M31 and M33 mass distributions. Regions are colour coded according to the mean M33 distance from the MW in kpc. 
}
\label{dtperi}
\end{figure*}

The probability for a close pericentre passage in the past and pericentre times for the standard mass model depends 
marginally on galaxy masses, being higher for larger M31 masses and lower for smaller M33 masses. If  the M33 mass does not vary with  time,  the probability of a pericentre passage   decreases from 18$\%$ to 16$\%$. A more sharp decrease in the probability is found if we consider galaxy mass growth with time from 9.2~Gyr ago to the present: in this case the probability of a close pericentre can drops from 18$\%$ to 6$\%$.

The pericentre distance increases as the velocity components of  M31 in the Galactocentric frame increase. 
There is a clear strong  anticorrelation of V$^{M31}_{x,y}$, the M31 x and y velocity components in the Galactocentric frame, with the M31 proper motion along RA. The mean separation at pericentre decreases  as the M31velocity along  RA increases (i.e. as M31 moves with higher speed towards the eastern direction).  Pericentre times decrease as the M31 velocity along DEC increase towards more positive values.

As an example we show in Figure~\ref{dtperi} the distribution of pericentre distances and times relative to four cases. In panel $a$ and $b$  the standard and low-mass models respectively have been used with the M33 mass that decreases from 9.2~Gyr to the present time, due to  mass removal during an M31-M33 encounter. Panel $c$ refers to the highvar mass model  in which galaxies  grow with time and today's M31 mass is high, in the range $1.8\times 10^{12} \le   M_{M31}  \le  3.2\times 10^{12}$~M$_\odot $. Clearly the probability of a pericentre in the past is higher if one considers a more massive halo for M31 but the mean pericentre distance does not vary much being D$_p$ of the order of 100~kpc of larger for most of the orbits. In all these 3 panels of Figure~\ref{dtperi} points are colour coded according to V$_{x}^{M31}$ that is strongly correlated with the velocity of M31 along RA. The closest pericentre distances are found for  negative values of V$_x^{M31}$ that correspond to small values of M31 transverse velocity along RA. Finally, in panel $d$ of Figure~\ref{dtperi} we show the distribution of pericentre distances and times when dispersions around the mean distances of M31 and M33 are considered for the standard mass model.
The colour bar clearly shows that M33  had a closer pericentre passage in the past around M31 if its distance today is less than 840~kpc.
If the M33 distance is indeed smaller than 840~kpc, disturbances in the outer disk are stronger because of the smaller pericentre distance, especially when M31 distance also increases  and M33 and M31 lie at a similar distance from the MW. However, as discussed in the next Section, we find that also for these cases the distribution of disturbed test-particles differs from that of a warp. We underline that  there is no evidence for the distance to M33 to be less than 840~kpc today.  The latest measurements using pulsating Cepheids do confirm with high precision the value of 840~kpc \citep{2023ApJ...951..118B} while the use of Mira variables, Tip of the Red Giant Branch, and RR Lyrae place M33 at a slightly larger distance than 840~Mpc and beyond M31 \citep{2023AJ....165..137O,2022ApJ...933..201L,2021MNRAS.508.3035S}.
Unless stated differently  we  consider distances fixed at their mean values in the rest of the paper.

\begin{figure} 
\includegraphics [width=9 cm]{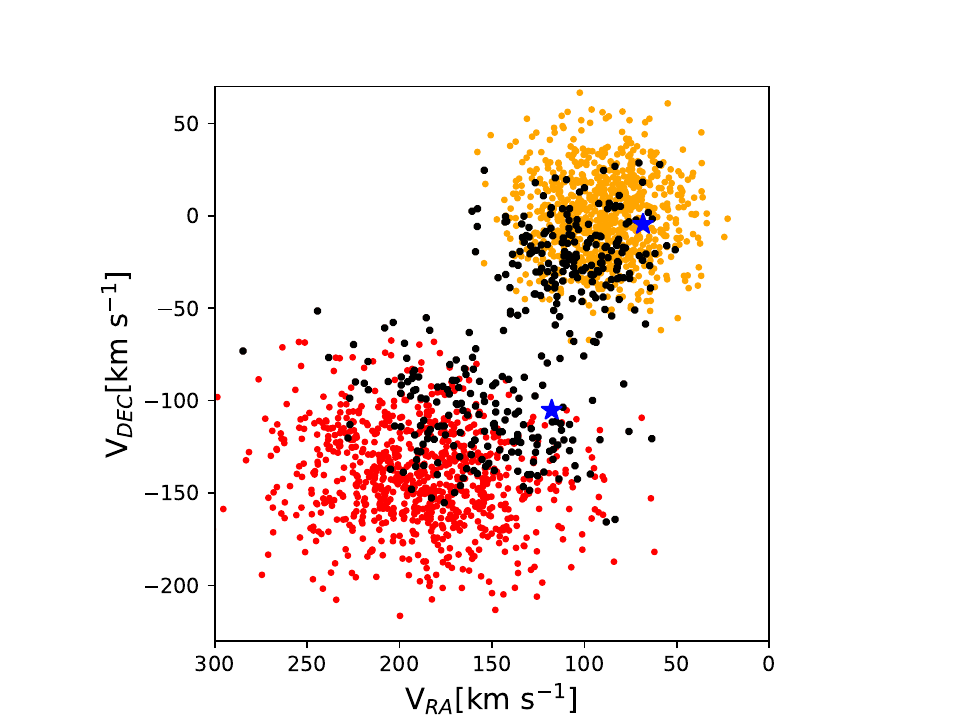}
 \caption{Components of the proper motion velocities along RA and DEC for M31 (red dots) and M33 (orange dots) sampling 1000 combinations of initial conditions. For each galaxy we circle in black the velocities for which we find a pericentre passage of M33 around M31 in the past (for the standard mass model).  The blue stars mark the values of the velocity components of M31 and M33 selected for showing the results of a test-particle simulation in the next Figure. For this case the MW, M31 and M33 mass at the present time are 1.35, 2.00, and 0.36 $\times$ 10$^{12}$~M$_\odot$ respectively. }
\label{periyes}
\end{figure}

To visualise which proper motion components favour a close encounter in the past between M31 and M33 we show in Figure~\ref{periyes} the distribution of the RA and DEC components of transverse velocities, V$_{RA}$ and V$_{DEC}$,  relative to M31 (red dots)  and to M33 (orange dots),  used to set the initial conditions of 1000 back orbit integrations. Masses  for MW, M31 and M33 have been sampled using the standard mass model for each orange and red velocity pair. Distances of M31 and M33 are fixed at their mean values. We circle in black the trasverse velocities   that results from orbits with a  pericentre passage in the last 9.2~Gyr. From the Figure it is clear that past pericentres are favoured by small values of  M31 V$_{RA}$ and V$_{DEC}$,   as well as  by negative values of  M33 V$_{DEC}$. The blue stars indicate the velocity pairs resulting from orbits that generate the M33 disk configuration shown in Figure~\ref{testpa}, as recovered from a test-particle simulation.

The general conclusion is that the probability for a close passage  of M33 around M31 in the past   for the standard masses of  Local Group spirals and {\it Gaia}-EDR3 is low but non-zero, and the average pericentre distance is rather large. In the rest of the paper we run test-particle simulations to study if  using any combination of
 galaxy masses and velocities, corresponding to orbits with a close pericentre passage of M33 around M31 in the past,  it is possible to reproduce  the M33 warp and its non  symmetrical distribution of dark  satellites.  
 
 \section{The M33 outskirts: Effects of a pericentre passage and the gas accretion scenario}
 
To investigate if some of the asymmetries observed in the M33 outer regions are related to interactions with its massive neighbour M31, we used  test-particles to model the disk of M33 and its dark satellites along the orbit using a wide range of initial parameters for the standard and low-mass mass models. The initial conditions for starting test-particle simulations 9.2~Gyr ago have been recovered through back orbit integration of the centre of mass of MW, M31 and M33.
 
 \subsection{The warp}
 
Given the range of galaxy masses and orbital velocities,  compatible with  {\it Gaia}-EDR3 and with dynamical models of rotation curves, as discussed earlier,  pericentre distances  are of the order of 100~kpc or larger, and we find  no excitation of a disk warp similar to that observed. The disk of M33 looses very few disk particles during a pericentre passage and gains  some dispersion that decays with time. There are initial conditions in positions and velocities for which the pericentre passage  triggers the formation of outer arms, rings and some mild warps only in the outermost 2~kpc region. Given these results, the rather low probability of a pericentre passage of M33 around M31, the large pericentre distance, and the fact that  this happened long time ago  we conclude that it is very unlikely that the disk warp observed in M33 is the result of a close passage around M31 in the past. 
 
\begin{figure*} 
\includegraphics [width=9 cm]{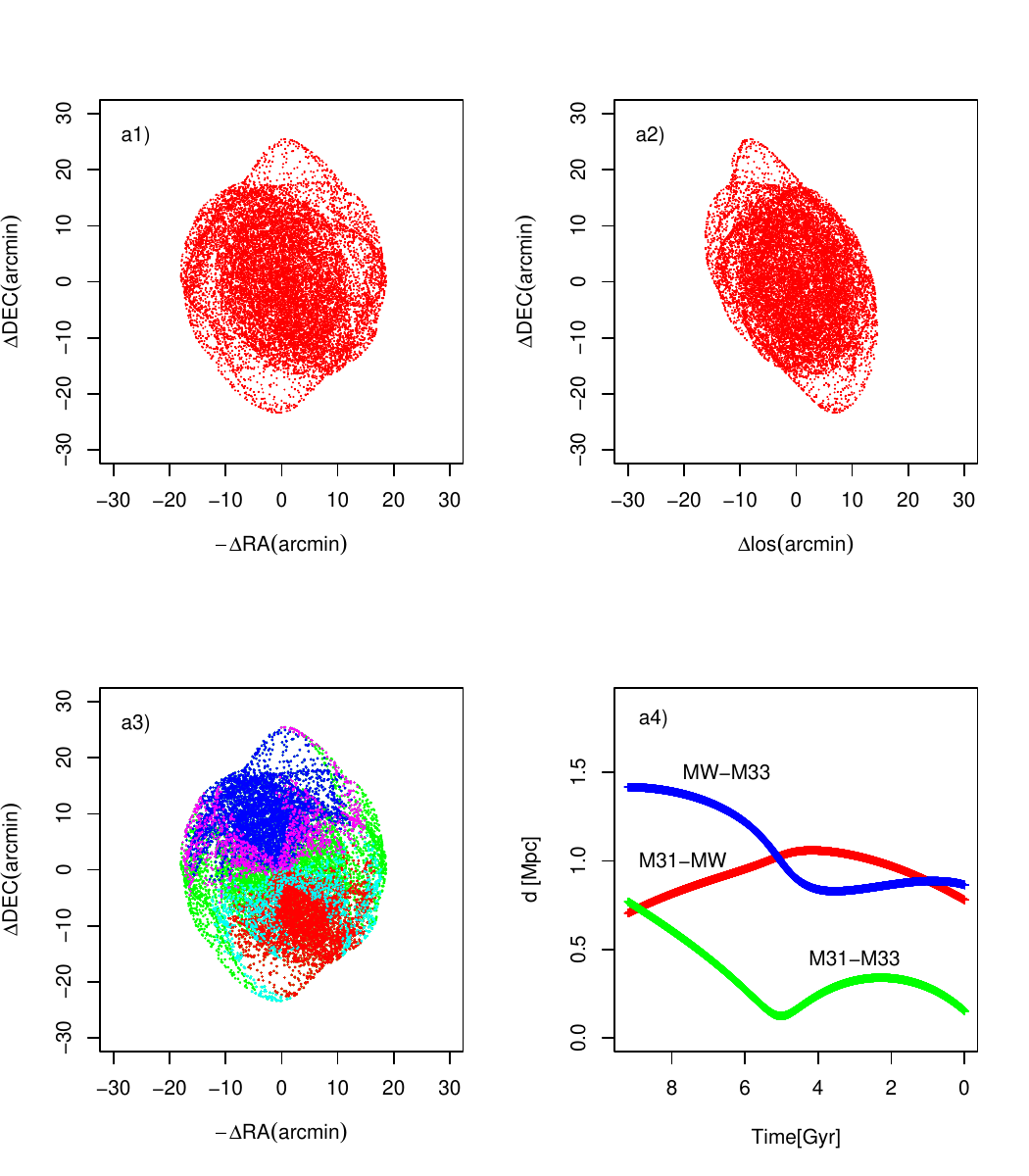}
\includegraphics [width=9 cm]{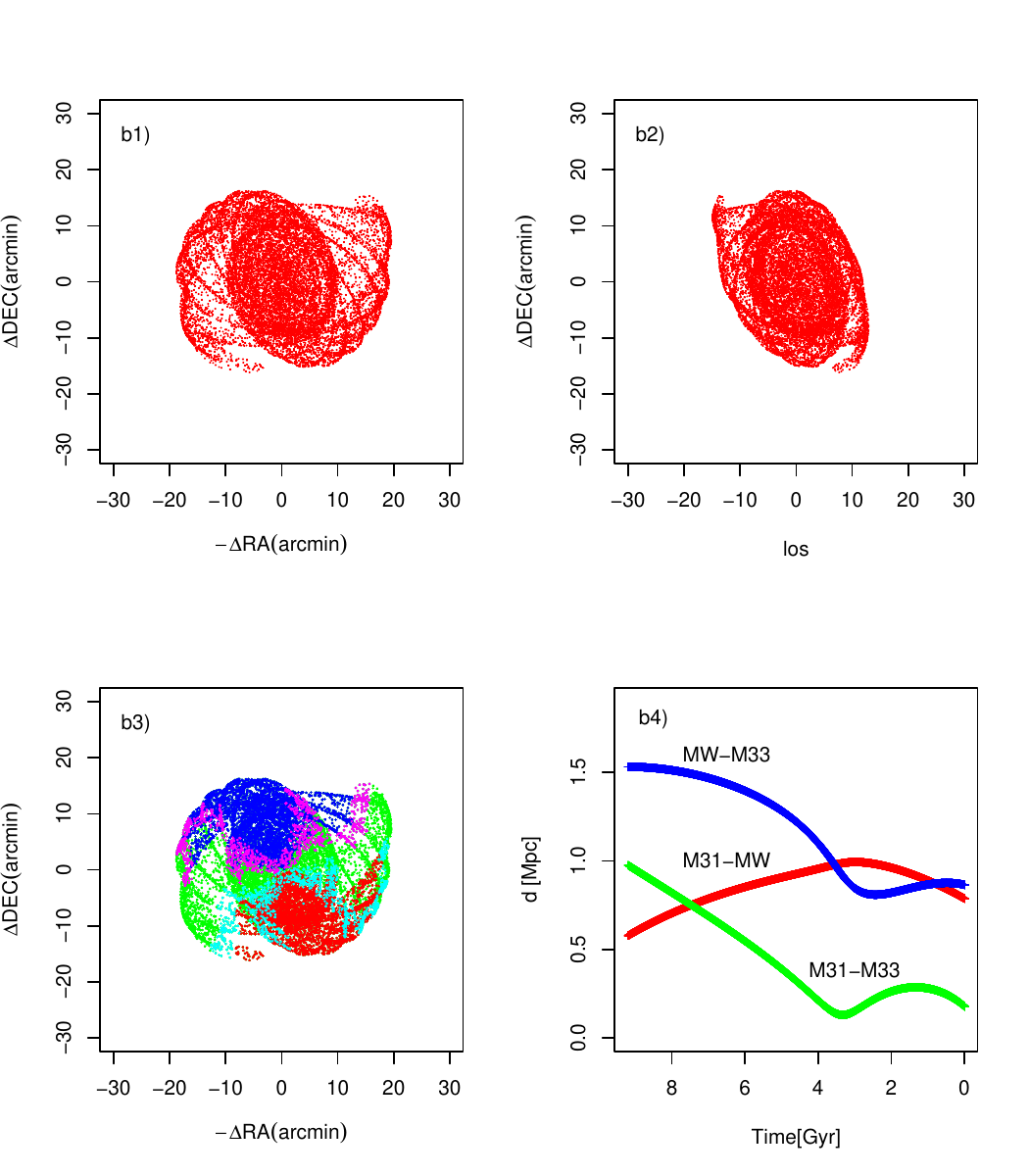}
\caption{ Space distribution and line of sight velocities of test-particles in the M33 disk at t=0 in the tangential frame and the time evolution of galaxy separations. They are results of a simulation that predicts tangential velocities shown by the blue star symbols in Figure~\ref{periyes}. Panels a) refer to a   standard mass model case,with the M31 mass equal to  1.5$\times 10^{12}$~M$_\odot$. The b) panels refer to a simulation for the highvar mass model, with M31 mass equal to  3.5$\times 10^{12}$~M$_\odot$. $\Delta$RA, $\Delta$DEC, and $\Delta$los are the  coordinate differences between the particle coordinates and the  M33 centre coordinates. Panels a3) and b3) show the line of sight velocities of the M33 disk, with the green colour indicating velocities close to systemic, blue and red the more extreme approaching and receding velocities respectively. In panels a4) and b4) the time evolution of the relative distances between MW, M31 and M33 in Mpc from 9.2~Gyr ago to the present time are shown. }
\label{testpa}
\end{figure*} 

As an example in panels a1), a2), a3), a4) of Figure~\ref{testpa} we show the results of M33 test-particle simulation for which the past pericentre passage of M33 around M31 generated a marked disturbance in the mass distribution, driving some weak outer arms along the M31-M33 direction.   The transverse velocities of the centre of mass of M31 and M33 resulting from their past orbits are shown in Figure~\ref{periyes} by the blue stars. For this case the MW, M31 and M33 mass at the present time are 1.35, 2.00, and 0.36 $\times$ 10$^{12}$~M$_\odot$ respectively. In Figure~\ref{testpa}  the distribution of particles with respect to the M33 centre in the RA-DEC plane and in the line of sight (los) - DEC plane are displayed. In the bottom panels of the Figure, we colour code the test-particles according to their line of sight velocity (with green colour indicating velocities close to the systemic) and the time evolution of the relative distance between the three galaxies (MW, M31, M33)  from 9.2~Gyr ago to the present time. The pericentre passage, for the initial conditions of the case shown, happened about 5~Gyr ago. If we now consider the same velocity vectors of M33 and M31 relative to the MW today, as for the case shown in panels a),  but for the highvar mass model with today's M31 mass equal to 3.0$\times 10^{12}$~M$_\odot$,  panels b) show the resulting morphology of the M33 disk, the line of sight velocities, and the time evolution of the relative distances.
A strong increase in the M33 mass clearly  makes the outer arms much more prominent. The pericentre passage of M33 around M31 in this case happened only 3~Gyr ago but there is no evidence that a warp similar to that observed is formed. 
By  sampling a few of these highvar  cases we noticed  that the stronger tidal disturbance generate prominent outer arms  which might change the orientation of the outermost isophotes adding a non negligible perturbation to the velocity field.

\begin{figure} 
\includegraphics [width=9 cm]{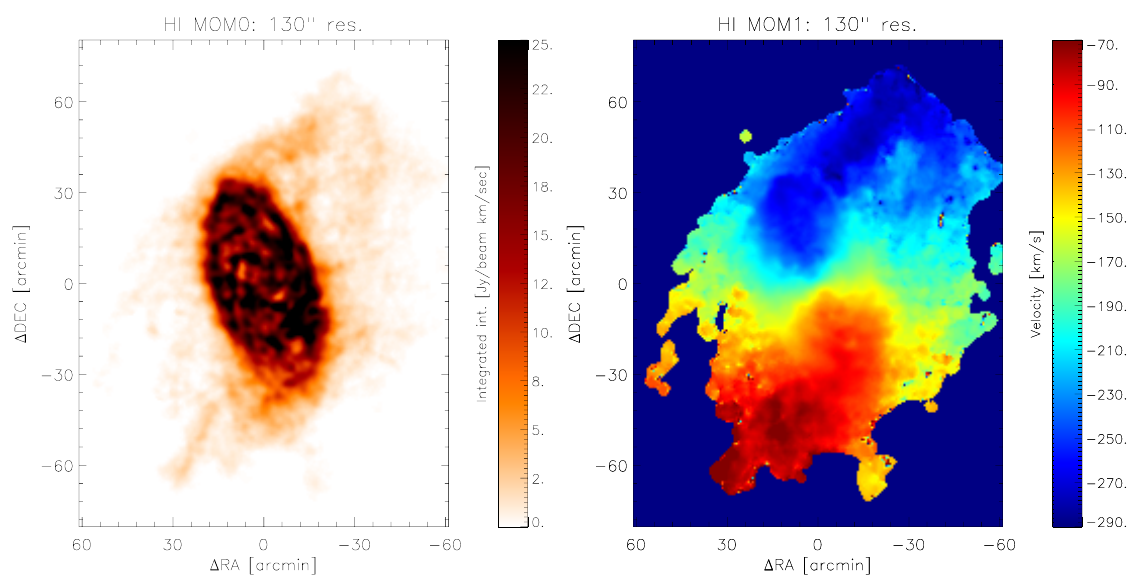}
 \caption{ HI intensity map and the intensity weighted mean velocity of M33 at  0.5~kpc resolution; see \citet{2014A&A...572A..23C} for more details. }
\label{HI}
\end{figure}

For comparison we show in Figure~\ref{HI} the HI intensity map and the intensity weighted mean velocity for M33 at  0.5~kpc resolution \citep{2014A&A...572A..23C}. Clearly the HI intensity map shows no sign of outer arms oriented anticlockwise and the velocity fields in the outer regions looks rather different than that shown in Figure~\ref{testpa}.

The excitation of a warp more similar to what is observed in M33 is possible under initial conditions that lead to centre of mass velocities now excluded by  proper motion measures. 
We furthermore exclude the possibility that the M33 warp is excited by a close bound massive satellite, possibly removed during a close passage around M31, because of the large pericentre distances and times inferred.

\subsection{A skewed distribution of gaseous satellites}

\begin{figure} 
\hspace{0 cm}
\includegraphics [width=9 cm]{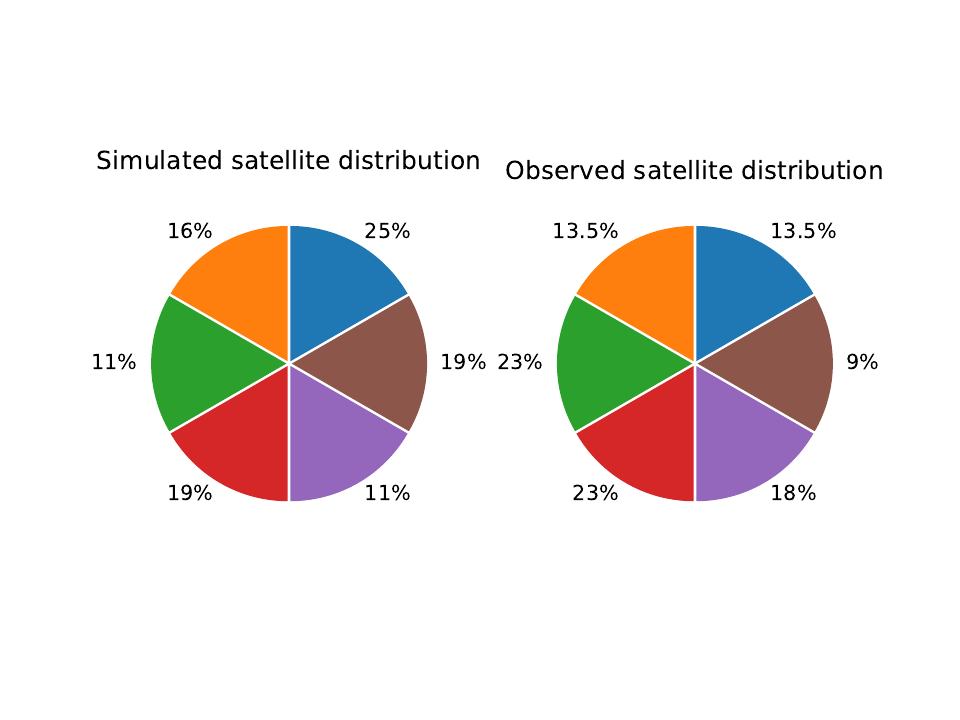}
 \caption{Comparison between the observed and simulated dark satellite distribution lying in projection beyond the outer HI disk of M33, averaged in 60$^\circ$ sectors. The observed distribution is from \citet{2008A&A...487..161G}. The  pie to the left summarises the average distribution of satellites in the various sectors for several  simulated orbits of M33 with a close pericentre passage around M31 in the past.  The percentages refer to the fraction of satellites observed or predicted  in each angular sector. }
\label{pie}
\end{figure} 

\begin{figure} 
\hspace{0. cm}
\includegraphics [width=10 cm]{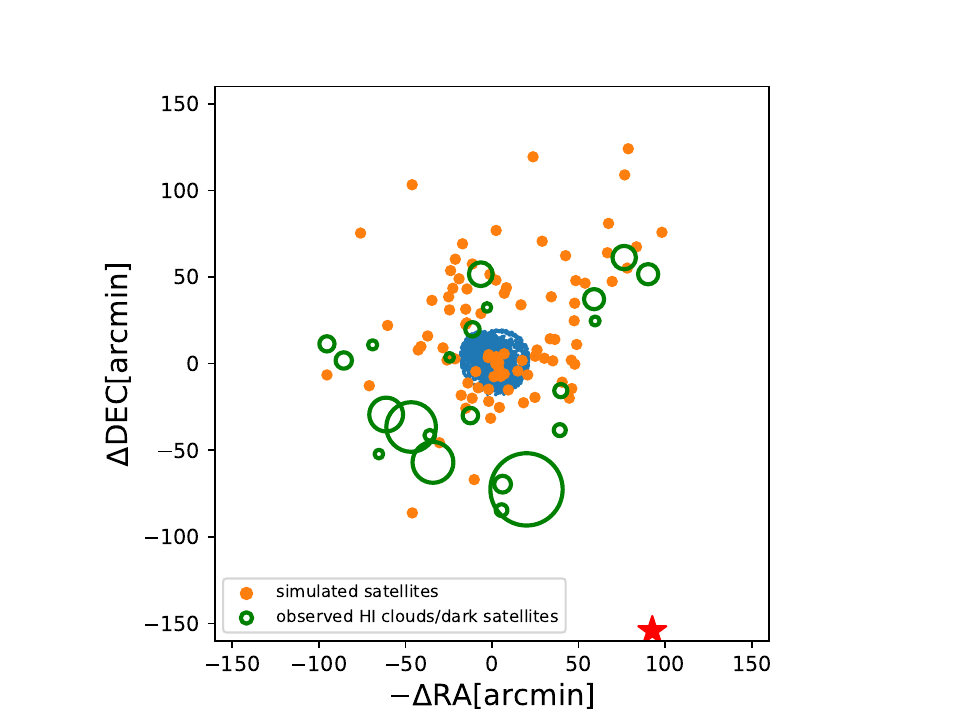}
 \caption{Example of one non-symmetrical satellites distribution around M33 induced by  a pericentre passage around M31. The initial conditions for the orbits are the same as those considered for the left panels of Figure ~\ref{testpa}. Test-particles in M33 disk are marked in blue, the simulated satellites are in orange, observed HI clouds (possible dark satellites of M33) are shown with open green circles whose sizes are proportional to  cloud masses. The red star indicates the position of AndiXXII.}
\label{sat}
\end{figure}

 As pointed out earlier, the  HI gas clouds in the M33 outskirts can trace the dark satellite population of M33  \citep{2008A&A...487..161G}.  If these were isotropically distributed around M33 about 9.2~Gyr ago, we can investigate if a pericentre passage of M33  around M31 had skew their spatial distribution making it  similar to that observed. The distribution of clouds seen in projection outside the lowest HI contour of M33  is in fact elongated in the direction of M31. In particular, there are fewer HI clouds (eight in number)  in the range -120$^\circ$ $\le$ PA$<$ 60$^\circ$  than in the southern region 60$^\circ$ $\le$ PA$<$ 180$^\circ$  and  -180$^\circ$ $\le$ PA$<$ - 120$^\circ$(14 in number) and these are also more massive ( PA being defined anticlockwise with respect to the celestial pole).    We ran test-particle simulations with satellites initially distributed isotropically around   some radial distance R$_{sat}$ from the M33 centre. In Figure~\ref{pie} we show the average angular distribution of the satellites resulting from several simulated orbits with a close pericentre  in the past, using  R$_{sat}$=50~kpc. For comparison the pie to the right of the Figure  shows the observed distribution of satellites found by \citet{2008A&A...487..161G} averaged  in 60~degree sectors around M33. We have not included in the statistics the faintest HI clouds discovered further out  by \citet{2012AJ....144...52L} towards M31. 

Figure~\ref{pie} clearly shows that although dark satellites of M33 are no longer homogeneously distributed after a close orbit around M31, the angular directions where the number of satellites is enhanced differs from the observed ones. Both simulated and observed distribution have an enhancement in the direction of M31 (30$^\circ$-90$^\circ$ sector), but the angular directions where very few satellites are found in the data are those around -120$\le$PA$<$60$^\circ$, while in the simulations satellites are lacking around PA=90$^\circ$ and PA=-150$^\circ$. 

In Figure~\ref{sat} we show the distorted distribution of satellites resulting from a close pericentre passage of M33 around M31 (orange dots), along the same orbit considered for Figure~(\ref{testpa}). The green open circles indicate the observed distribution of satellites if these are associated with HI clouds. The circle size is proportional to the cloud HI mass.  The largerer HI masses observed for satellites in the south-east further increases the discrepancy with the simulated distribution if the observed large masses are due to an ensamble of low mass units,  unresolved by the large beam of the Arecibo radio telescope. In the Figure we mark with a red star symbol the location AndXXII, one of the two optical satellites that might be bound to M33, the other one lying at larger projected distance \citep{ 2013MNRAS.430...37C,2022MNRAS.509...16M}. Concerning the missing satellite problem for M33 we underline that satellites can be removed by a close pericentre passage around M31 if these were originally located at radial distances larger than about 50~kpc. This is in agreement with the explicit computation by \citet{2018MNRAS.480.1883P} that show how the tidal radius of M33 varies as a function of the pericentre distance.  We estimated for example that roughly  30$\%$ of the satellites located at 70~kpc radial distances from M33 centre can be lost in space after a pericentre passage around M31. Test-particle simulations have shown that in this case the satellites  removed from their orbit during a close passage of M33 around M31, should be found in the direction opposite to M31 at very large radii, beyond 100~kpc. This is because they are lost in space where the encounter took place in the past.

\subsection{Condensations from an intergalactic filament}  

The HI warp and  the skewness of HI cloud distribution in M33 outskirts can result from gas condensations accreting into a galaxy along a preferential plane as due for example to the presence of an intergalactic filament.  We may expect the baryonic gas mass to assemble first chaotically, but later gas vorticity inside intergalactic filaments can drives slow gas and dark matter accretion. Gaseous perturbations might condense retaining their angular momentum and settle into galaxies  along a plane related to the filament vorticity   \citep{2009ApJ...700L...1K,2009ApJ...694..396B,2010MNRAS.408..783R,2020MNRAS.495L..42R,2023ApJ...944..143W}. Numerical simulations and dedicated observations have shown in fact that cosmic filaments have spins and vortical flows \citep{2021NatAs...5..839W,2021MNRAS.506.1059X}.
Although streams can happens from different directions, the net result is a preferential orientation of the angular momentum of the accreted gas and dark matter. There is in fact evidence, from both observations and hydrodynamical simulations, that  galaxy's spin are not randomly oriented with respect to the filament's geometry  \citep[e.g.][and references therein]{2018MNRAS.481.4753C,2022MNRAS.516.3569B}. The halo spin of galaxy disks tends to be oriented along the longest filament dimension  due to the filament's  gravitational pull.  Massive ellipticals and disks with large bulges, experiencing  numerous merging, flip their spin perpendicular to the filament direction. There is however some discrepant findings both on the theoretical and observational side concerning the spin-filament allignment for intermediate and low-mass galaxies in the Local Universe \citep{2018MNRAS.481.4753C,2019ApJ...876...52K}. Some observations underline  that disks in less massive halos with no massive bulges, such as M33,  retain  memory of the  filament anisotropic environment  \citep{2022MNRAS.516.3569B}. 

Following these suggestions we considered the possibility that the angular momentum of the warped outer disk of M33 is in close alignment with  the local filament direction. If a cosmic filament is embracing M31 and M33,  with M33 being accreted more recently by the local filament after the formation of the inner disk,  the outer disk spin might reflect that of the filament.
Given the inclination, rotation and PA of the M33 outer disk, its spin points away from us towards the south-western direction, forming an angle of about 50$^\circ$ with the line of sight (the inner disk spin forms an angle of about 50$^\circ$ with the line of sight, pointing away from us towards the north-western direction). Both the inner and  outer disk spin of M31 are instead oriented towards us,  with the inner disk spin  forming 77$^\circ$ with the line of sight, at PA$\simeq 130^\circ$ \citep{2010A&A...511A..89C}, close  to the M33 direction (PA$\simeq 137^\circ$ \citep{1997ApJ...479..244C}). This would be consistent with the picture that M31 is undergoing a spin flip due to recent merger events, with its specific star formation rate being low because of the lack of  gas accretion from the filament.

The outer disk of M33 has a total gas mass of the order of 5$\times 10^8$~M$_\odot$ \citep{1997ApJ...479..244C} and   a sparse population of young stars with ages 100-200 Myr  mixed with a faint old stellar population \citep{2011A&A...533A..91G}. Since the outer disk HI surface density  is below 1~M$_\odot$~yr$^{-1}$  most likely star formation in the outer disk happens in bursts related to  gas accretion events. These  increase and compress the gas surface density  triggering star formation.  If the mean inflow rate from the cosmic filament is of the order 1~M$_\odot$~yr$^{-1}$, as estimated from possible gas infall from the observed HI clouds  \citep{2008A&A...487..161G},  this would be sufficient to fuel the inner disk star formation rate of 0.5~M$_\odot$~yr$^{-1}$ \citep{2009A&A...493..453V} and at the same time sustain the formation of the gaseous and stellar outer disk through the last few Gyr.   The evidence of ionised gas inflow towards the M33 disk \citep{2017ApJ...834..179Z} is for now restricted to metal rich gas, likely  related to a galactic fountain, but  future radio and optical telescopes with high resolution and sensitivity can further support this scenario.

\section{Summary and conclusions}
 
Our Local Group, hosting the MW, M31, and M33 in addition to numerous dwarfs, provides the nearest laboratory for studying the dynamics and evolution of galaxies. In this paper we have used the detailed knowledge of  M31 and M33 halo properties, together with results of deep observations of  M33 outskirts and the latest measurements of the M31 and M33 transverse motion, to shed light  on the  role of the M31-M33 past interaction  for shaping the M33 outer disk and satellite distribution.  Transverse motion measurements, that are only possible today for Local Group galaxies, allowed us to reconstruct the galaxy orbital history. We have used {\it Gaia}-EDR3  \citep{2021MNRAS.507.2592S} to integrate backwards in time for 9.2~Gyr the centre of the mass motion of the three Local Group spirals for  a  range of halo masses and velocities compatible with  rotation curves and proper motion data.  In the forward integration, we used test-particle simulations to investigate if a close interaction between M31 and M33 in the past can explain the observed M33 warp and the dark satellite distribution.  

N-body simulations have been used to evaluate possible dark matter losses and the dynamical friction term. By tuning test-particle simulations with N-body simulations, we  find that  the analytical expression of the  Coulomb logarithm can be approximated by the  ratio between the distance of the secondary from the primary and the scale length of the halo, that is to say $C=1$ for the orbital solutions and the masses we examined. We also determined  mass losses of M33 along orbits that brought M33 close to M31 in the past.   For these cases we find that the mass of the M33 dark matter halo decreases by 35-40$\%$. 

By considering Gaussian or random distributions in the parameter space,  we evaluated the probability for M33 to be on its first infall to M31.   We find that this probability is higher than 70$\%$ using the most quoted values of  Local Group spiral galaxy masses, but a  pericentre passage in the past cannot be excluded. 
We considered both the case of today's M33 dark matter halo mass equal  to that determined by the fit to the rotation curve,  or a lower halo mass   resulting from mass losses during the tidal encounter with M31. In this case the probability of a close preicentre passage around M31 in the past is as low as 11$\%$. For  both cases  pericentre distances  are of the order of  100~kpc or higher. By examining a wide range of initial conditions, we conclude that if an encounter between  M31 and M33 took place in the past,  outer disk disturbances would be    mild because of the large pericentre distance. Given the observed transverse velocity components of M31 and M33, planar outer rings or spiral arms   possibly form and survive rather than a warp similar to what has been observed in M33.
We also examined orbits for  some extremely high values for the M31 mass, suggested by  a recent analysis of satellites orbital angular momenta \citep{2023ApJ...948..104P}, and considered that galaxies grow in mass with time. Although, for given values of galaxy masses at t=0, mass growth decreases the past pericentre probability,  the increase in the M31 mass makes the probability for a close encounter with M33 in the past close to 50$\%$.   However, the formation of a warp, similar in amplitude and orientation to what has been observed, does not seem compatible with the transverse motion inferred by Gaia-EDR3.

Our results on the M33 warp and the poorly convincing match between the skewed observed dark satellite distribution (counterpart of observed HI clouds around M33)  and the simulated one,  excludes the possibility that a past closer encounter between M31 and M33 played a role in shaping the outskirts of M33. We have outlined a possible scenario of slow gas accretion from cosmic filaments to explain disk warps as that observed in M33.   In this case the HI clouds can be interpreted as the cold baryonic component of  matter condensations from cosmic filaments falling into the galaxy. Although the expected mass flow rates are consistent with M33 star formation rate, and  this scenario predicts a different orientation of the M31 spin due to merger events, further investigations are needed to confirm  the link between gas accretion from cosmic filaments and  outer disk misalignment in late-type disk galaxies such as M33.  

 M33 has a low baryonic fraction, of the order of 0.02 \citep{2014A&A...572A..23C}, that is much lower than the expected cosmic value. This low value, consistent with the baryonic to halo mass relation found from abundance matching for galaxies of a similar mass \citep{2013ApJ...762L..31B, 2019A&A...626A..56P}, requires that baryons have been blown out by supernova feedback. However, hydrodynamical simulations struggle to produce  systems that expel a lot of gas given their  inferred rate of star formation \citep[e.g.][and references therein]{2019MNRAS.485.2511T}. If today M33 is not on its first infall into M31 and if the dark matter halo mass  was larger in the past, this would make the problem of baryonic losses more challenging, also for models that  include radiative feedback  \citep{2014MNRAS.445..581H}. Placing some of the missing baryons in an extended halo is an alternative possibility  \citep{2019MNRAS.485.2511T}, that needs to be explored observationally in the future given the uncertain presence of a baryonic halo for this galaxy \citep{2013MNRAS.428.1248C,2020MNRAS.499.4716C}.

\begin{acknowledgements}
 We would like to acknowledge the anonymous referee for his/her comments to the original version of the manuscript. EC acknowledges financial support from  INAF-Mini Grant RF-2023.
\end{acknowledgements}

\begin{appendix}
\section{The M33-M31 orbital evolution: Models and results of previous studies}

Computation of the M33 and M31 orbits and the effects of a possible close encounter between these galaxies has been carried out in the past. 
  
Some papers have addressed the question of a close encounter between M31 and M33 independently on the initial conditions that bring them into the actual configuration (sky projection and proper motion). Other papers have instead studied the past and future orbital history of these galaxies and of the MW with the goal of reproducing accurately their position and velocities as observed from the solar system. There is not yet an analysis of the M33 matter distribution resulting from orbits that are consistent with the M31 and M33 proper motion range as inferred by the {\it Gaia} satellite data. We underline to this purpose, that in order to select orbits which bring M33 and M31 in their actual sky position, and to verify the M33 warp orientation with respect to our line of sight, it is necessary to consider the MW potential well because of the mutual attraction between MW and  M31. We summarise  below the main assumptions and numerical approaches used for studies concerning the possible M31 and M33 past interaction.

\citet{2005ApJ...633..894L}, \citet{2008MNRAS.390L..24B} and \citet{2009Natur.461...66M} have used test-particle simulations for the M33 disk or N-body simulations to study  disturbances induced by a past encounter with M31. 
However, the absence of any measurement of M31  proper motion, at the time these papers have been published limits the validity of their results. In addition a few assumptions made  need to be revised such as those concerning  the MW mass, considered a factor 1.5 higher than the M31 mass \citep{2008MNRAS.390L..24B}, the underestimate of the M33 mass (3.8$\times$ 10$^{10}$~M$_\odot$ in \citet{2008MNRAS.390L..24B}), the simplified  matter potentials without considering dynamical friction effects. These studies conclude that a warp develops in the disk of M33 after a pericentre passage happening less than 1~Gyr ago.
A similar result has been also reached by  \citet{2009Natur.461...66M} who did not run test-particle simulations but  analysed only the centre of mass motion estimating  the likelihood of pericentre passage. No constraints were available for M31 proper motion at the time this paper  has been published.

Data on M31 transverse motion has been first analysed and presented by \citet{2012ApJ...753....7S,2012ApJ...753....8V} using HST data.  In these papers the mass of M33 has been correctly considered non negligible compared to M31 mass. The paper focus, however,  was the fate of Local Group spirals and not the role of galaxy encounters  in shaping  galaxy's outskirts, likely because  of the  large pericentre distances they found. Using the data of  \citet{2012ApJ...753....7S,2012ApJ...753....8V}  on M31 transverse motion and statistics from the Illustris cosmological simulation,  \citet{2017MNRAS.464.3825P} analysed the past orbits of M33 and M31 and conclude that  it is very unlikely that M33 made a close pericentre passage around M31 in the past.  In searching for Local Group analogues in cosmological simulations the M33 mass has however been underestimated.

 \citet{2016MNRAS.456.4432S}  estimated the velocity vector of M31 by modelling the galaxy and its satellites as a system with cosmologically motivated velocity dispersion and density profiles.  The cosmological simulations used did not include however, the baryonic physics which can impact on the abundance and properties of dark matter substructures.
The resulting radial velocity is consistent with the observed one by \citet{2012ApJ...753....7S,2012ApJ...753....8V}, while the tangential component is much higher. A hydrodynamical model of M33 and of its interaction with M31 has been thereafter presented by \citet{2018ApJ...864...34S} based on these tangential velocities with the purpose to reproduce the M33 gaseous warp 
 and the spiral pattern as induced by tidal interaction with M31. Using the backward and forward orbit  integration scheme  \citet{2018ApJ...864...34S} do not find orbits  potentially able to explain the disturbances around M33.  However, considering some variations on the initial conditions  estimated by  \citet{2016MNRAS.456.4432S} and following an orbital solution that predicts a pericentre distance a of 37~kpc happening less than 2~Gyr ago they simulated the M33-M31 interaction using an N-body/hydrodinamical simulation.  They conclude that the gaseous warp cannot be reproduced to a satisfying degree of similarity given the presence of strong outer spiral arm in the simulated gaseous distribution. In the same year a paper by \citet{2018MNRAS.478.3793D} showed that the non symmetrical spiral pattern might result from star formation feedback coupled to gravitational instabilities in the stellar and gaseous disk.
  
The effects of a close pericentre passage on the satellite distribution of M33 has been analysed by \citet{2018MNRAS.480.1883P}. M33 satellites considered in this work are those with total mass above the minimum mass for star formation to occur (roughly 10$^{7.4}$~M$^\odot$). These authors agree that it is likely that M33 is on its first infall towards M31 or had a wide pericentric passage in the past. In this case its satellites are expected to remain bound and the authors
favour deep searches around M33. The discover of a second satellite of M33 \citep{2022MNRAS.509...16M,2023arXiv230513966C} alleviates the tension concerning the missing satellite problem for this galaxy.
Both satellites however need to be more massive than currently estimated and need to much closer (distances of a few kpc) to induce the warp  in M33 \citep{2018ApJ...864...34S}. This close distance implies that the M33 inner disk would be {\it perturbed} while it seems quite undisturbed. 

The {\it Gaia} satellite data  \citep{2019ApJ...872...24V,2021MNRAS.507.2592S} has refined the M31 proper motion measurement of  \citet{2012ApJ...753....7S,2012ApJ...753....8V} and strengthen the conclusion that M33 is on its first infall into M31 although no searches has been done in the allowed parameter space of mass and motion of the Local Group spirals.  
\citet{2020MNRAS.493.5636T}  analysed the orbital history of M33 using the best values of  {\it Gaia}-DR2 proper motion measurement, accounting for dynamical friction and mass losses experienced by M33 in the semi-analytic back orbit integration.  A full N-body/hydrodynamical simulations  has been used to follow the orbits forwards in time.  The authors find a close pericentre passage with D$_p$=50~kpc which excited the warp in the outskirts of the extended baryonic distribution of M33.  This warp survives until now although the close encounter happened 7-8~Gyrs ago. However, we underline in what  follows some limitation of their study: a) The MW perturbation  on the M31 orbit has not been not considered; this limits also the ability of their approach to reproduce the exact position of the galaxies on the plane of the sky and the warp orientation with respect to our line of sight, b) The M33 mass  has been underestimated, c) A very large value for the Coulomb logarithm has been used in the dynamical friction formula possibly due to the fact that they assumed dark matter distributed in a Plummer sphere. This can increase the dynamical friction term by a factor 1.5-2 \citep{2008RMxAC..34...83E} for a given halo scale length. Moreover, the M33 Plummer  scale length comes out smaller by a factor four than the equivalent  Hernquist halo scale length (of the order of 50-60~kpc for M33, given the fitted NFW profile to the observed rotation curve). And finally  they found that mass losses further increase the Coulomb logarithm by a factor five.  The end results is that the dynamical friction term considered is a factor ten higher than what can be inferred using a Hernquist model for the dark halo potential, 
d) They assumed a disk extension for M31 and M33 larger than  40 kpc. This is not supported by today's observations of M33 that give an extension of the star forming disk of only  8.5 kpc, with the outer fainter disk reaching  about 20~kpc. Furthermore, there is no evidence of an HI gas tail they predict in the direction opposite to M31  from HI data \citep{2004A&A...417..421B}.
e)The exact values of the {\it Gaia}-DR2 proper motion measurements for M31 has been used; these are well outside 1-$\sigma$ values of {\it Gaia}-EDR3 measurements.   {\it Gaia}-EDR3  agree with HST data, not considered by the authors, thus casting doubts on the predicted mall M31-M33 pericentre distance in the past.

\section{ The N-body simulations used for determining reliable orbits in test-particle simulations}

Test-particle simulations include  dynamical friction  and  mass losses during the M31-M33 close encounter.  The parametrisation and coefficients of these effects needs to be determined by matching the results of these simulations with N-body simulations. The  analytical expression  of the dynamical friction induced by a primary galaxy onto a secondary galaxy (see Eq.(7))  has a Coulomb factor  parametrised as in Eq.(8)  following  \citet{2012ApJ...753....9V} and \citet{2003ApJ...582..196H}.  The coefficient C has been determined to be 0.71 by \citet{2003ApJ...582..196H} or 0.82 by \citet{2012ApJ...753....9V} using N-body simulations  to calibrate semi-analytic predictions. In our case, however, the mass ratio between the primary and secondary galaxy is different than considered in previous papers (closer to 1:3 rather than 1:10) and we also take into account the friction exerted by the secondary over the primary galaxy, and possible variations of galaxy masses after a pericentre passage.  Moreover, we don't simulate the final merged state between the two galaxies, as in the two quoted papers, but stop the simulation when the M31-M33 position is what we observe today. Therefore it is unclear which value of C better approximate dynamical friction along the orbits of interest to this study.

We used  N-body simulations  based on hierarchical tree methods and fully vectorised \citep{1987ApJS...64..715H}. These simulations  are adequate to  describe collisionless systems such as stellar disk and dark matter halos and have often been used to simulate galaxy encounters \citep[e.g.][]{1992ARA&A..30..705B}.  The stellar disk of M33 extends as far out as the whole gaseous warped disk of M33 \citep{2011A&A...533A..91G} and the rotation curve of this galaxy clearly indicate that an extended dark matter halo is in place. Gas has not been explicitly considered because hydrodynamic effects are not addressed in this paper. To simulate M31 we distributed the particles in a sphere following a Hernquist density profile. This has a finite mass equal to the total mass of M31 (baryonic + dark matter). Given the total mass and the dark matter concentration parameter c$_{vir}$ of the NFW profile fitted to the rotation curve data, we find the parameters of the  Hernquist profile that has the same enclosed mass within r$_{vir}$ \citep{2012ApJ...753....8V}. Similarly, to simulate M33 we distributed most of the particles in a sphere with a Hernquist density profile, and mass equal to the dark mass of M33.  A small fraction of the particles have been used to simulate the M33 disk that has a total mass equal to the sum of the stellar and gas mass of M33 as given by \citet{2014A&A...572A..23C}. This disk has an exponential profile and extends out to about 20~kpc. 
Being the closest distance between M33 and M31 much larger than the M31 disk, for all the orbits considered, we did not distribute the stellar mass of M31 in a disk, a bulge and a halo component but considered a halo with mass equal to the sum of the baryonic and dark matter. 
The typical  number of particles used to simulate each galaxy, M31 and M33, is 40,000, while only one particle is used to simulate the MW and its gravitational potential. The dark matter particles and star particles in M33 are  roughly of equal mass using a  number of particles of the order of 1000-1500 to simulate the M33 disk, given the low baryonic fraction of this galaxy. 
 
\begin{figure} 
\vspace{-2.5 cm}
\includegraphics [width=9 cm]{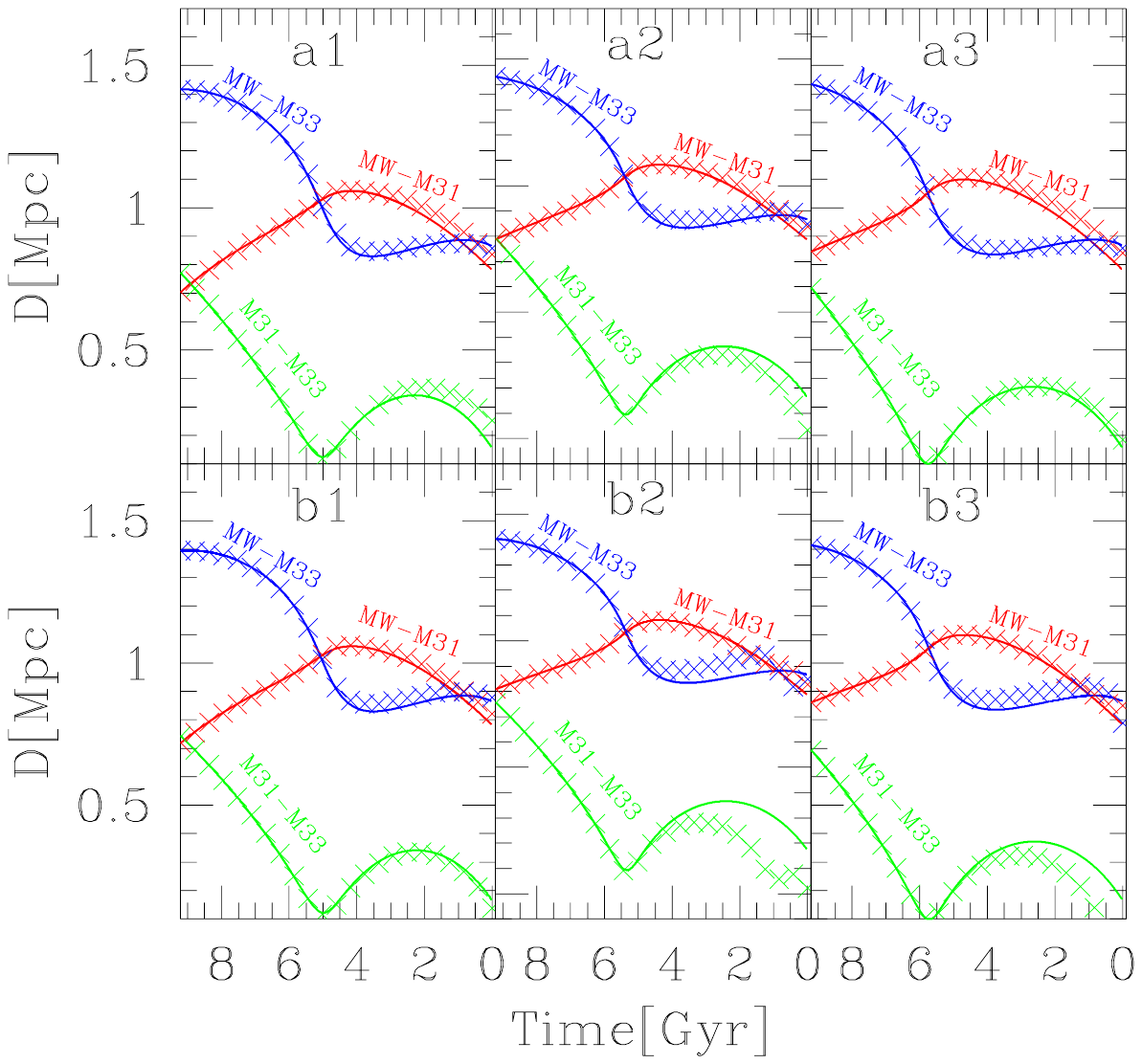}
\vspace{-2. cm}
 \caption{Time evolution of galaxy separations for three different sets of initial conditions at t=9.2 (today): the continuous lines in panels a1, a2, and a3 refer to the  results of  test-particle simulations for C=1.0; panels b1, b2, and b3 are  for the same  initial conditions but for  C=0.88. Crosses in each panel indicate galaxy separations resulting from N-body simulations with the same initial conditions as  test-particle simulations.}
\label{b1}
\end{figure} 

The initial conditions used for N-body simulations are recovered through the backward integration of the MW, M31 and M33 centre of mass in the Galactocentric frame, as described in Section~4. We ran several N-body simulations and test-particle simulations to find the value of C for which the orbits described by  test-particle simulations matches those recovered by N-body simulations. We started with initial conditions retrieved by test-particle simulations using C=0.6. Using N-body simulations we found that most of the orbits with a past encounter have an M31-M33 distance today  shorter than what is observed or the two galaxies have already merged. Using C=1.2 in the test-particle simulations instead the results of N-body code  indicated the opposite: orbits with a past encounter take M33 today at a larger distance from M31 than it is actually observed. The overall agreement between  test-particle  and N-body simulations is very good for C$\simeq 0.85 -1.05$. Within this interval  we have that for about half of the initial conditions leading to a close encounter between M31 and M33 in the past, M33 experienced a stronger dynamical friction and it is slightly closer to M31 than it is observed. The opposite is true for the other half of initial conditions. The overall agreement between the orbits from test-particle and N-body simulations is very good, as shown by plotting the time evolution of galaxy separations  in Figure~\ref{b1} for a few examples.  The continuum lines in the Figure show the time evolution of the distances between MW-M33 (blue lines), MW-M31 (red lines), M31-M33 (green lines) for three different sets of initial conditions. We have used C=1 to recover the initial conditions with the backward integration scheme for the runs in the upper panels and  C=0.88 for the same cases in the bottom panels. The crosses show the results of the corresponding N-body simulations. As we can see there is not a value of C for which the agreement is better for any set of initial conditions. Small fluctuations are present with orbits  that have slightly faster or slower decay. The correspondence seems slightly better for C=1 and we adopted this value for the study presented in this paper.

\begin{figure} 
\vspace{-1.3 cm}
\includegraphics [width=9 cm]{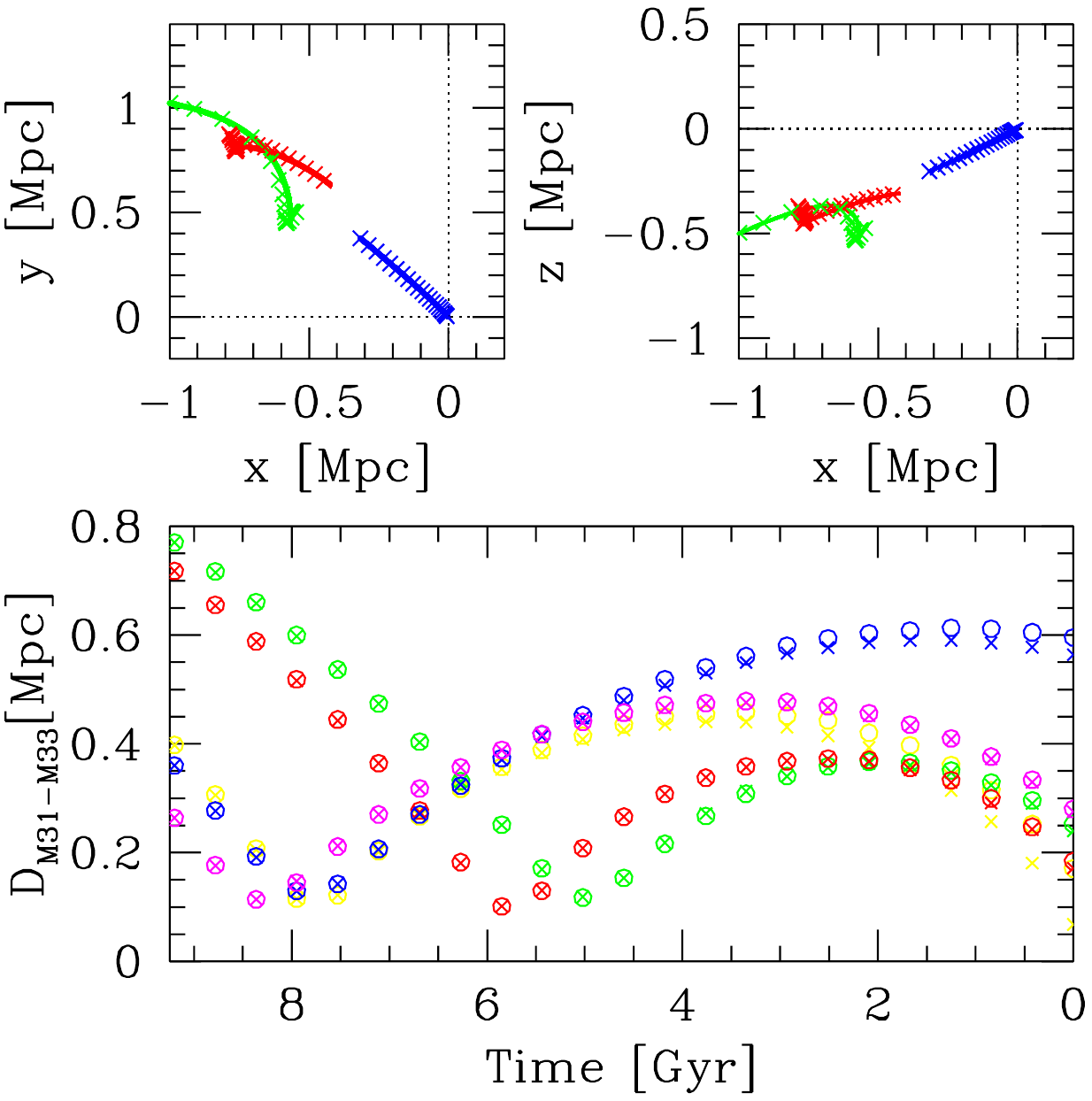}
\vspace{-2. cm}
 \caption{Comparison between results of test-particle and N-body simulations for the orbital evolution of Local Group spirals. Upper panels: Orbital evolution of the centre of mass of MW (blue), M31(red), and M33(green) for a set of initial conditions. Continuous lines are for test-particle simulations with C=1 and crosses are the results from the corresponding N-body simulations. Bottom panel: Time evolution of the M31-M33 separation as given by N-body simulations for five sets of initial conditions (corresponding to the five colours) with C = 1.0. The open circles are for simulations that used 10000 particles  for each galaxy while crosses refer to simulations with 40000 particles for each galaxy. }
\label{b2}
\end{figure} 

In the upper panel of Figure~\ref{b2} we show also the good match between the MW, M31 and M33 orbits resulting from a test-particle run with C=1 (continuous lines) and the corresponding N-body simulation (crosses). We also checked the convergence of N-body code by varying the number of particles from 10000 to 40000 for each  halo and for several initial conditions. For most cases there were no significant differences in the orbits considering 10000 or a higher number of particles. As an example, in the bottom panel of Figure~\ref{b2} we show the time evolution of the centre of mass separation between M31 and M33 for 10000 and 40000  halo particles and  five sets of initial conditions recovered using C=1. 

\begin{figure} 
\vspace{-5.8 cm}
\includegraphics [width= 9 cm]{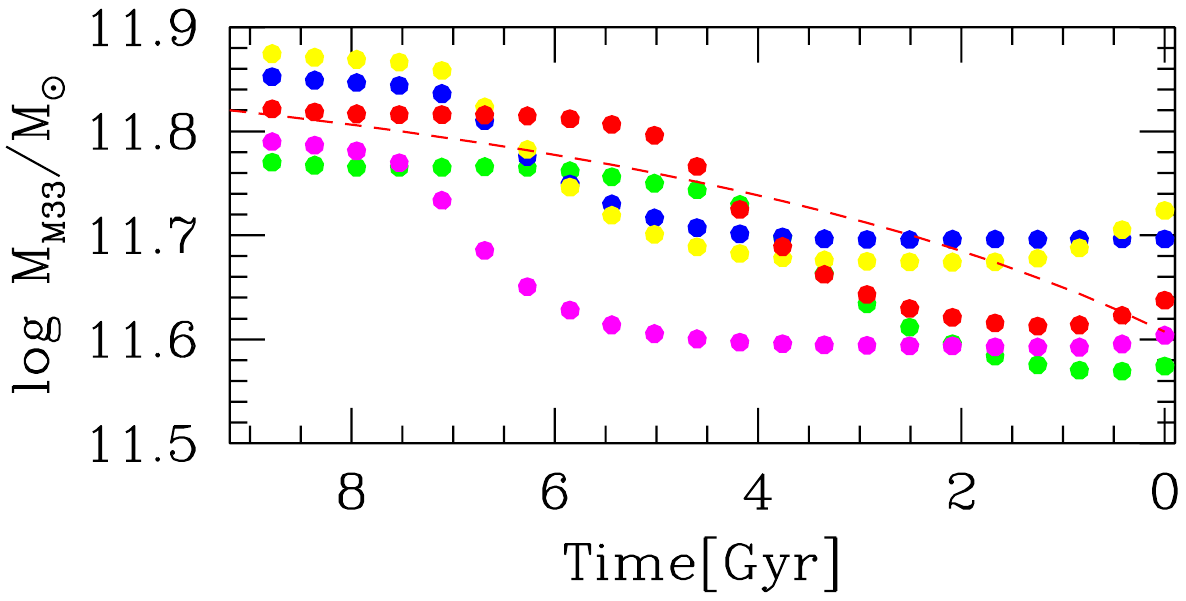}
\vspace{-2. cm}
 \caption{Results of N-body simulations for the time evolution of the M33 mass along five orbits (one for each colour) with a close pericentre passage of M33 around M31 in the past. The initial conditions and colour coding are the same as for cases shown in the bottom panel of Figure~\ref{b2}. The red dashed line shows for comparison the analytical expression used in the semi-analytical computation to describe time variations of the M33 mass for the case shown by the red filled dots.}
\label{b3}
\end{figure}

We underline that the agreement between  N-body and  test-particle simulations has been found considering also mass losses from the M33 halo after a close pericentre distance. If we neglect mass losses in test-particle simulations the required value of C is only slightly smaller, that confirms the earlier value found by \citet{2012ApJ...753....9V},  where mass losses have been neglected. We  display in the left panel Figure~\ref{b3} the decrease in mass  of the M33 halo along the orbit for the five cases shown in Figure~\ref{b2}. Mass losses for M33 depends marginally on initial conditions and have an average value of 38$\%$ for orbits with a close pericentre passage of M33 around M31 in the past. These mass losses have been computed by considering the dark matter mass inside the virial radius for the Hernquist profile. We recall that the M33 and M31 dark matter virial radii are much larger than 100~kpc (300~kpc and 190~kpc for the M31 and M33 halo respectively for masses at the central value of the range considered for the standard model). Therefore the mass inside the M33 viral radius might slightly increase as the galaxy reaches today's configuration, due to halo crossing, with the increase being more evident for orbits that predict a smaller M31-M33 separation.

The mass-loss term considered in our semi-analytical model is a good approximation given the mass losses inferred through N-body simulations. The analytical expression of the rate of mass loss used, however, predicts more steady losses while during the galaxy-galaxy encounter most of the mass is lost close to pericentre, as Figure~\ref{b3} shows. Because there are no estimates of the pericentre time for any given set of initial conditions, it is not easy to improve further the mass-loss model.

\end{appendix}

\end{document}